\documentclass{article}
\pdfoutput=1
\usepackage{graphicx}
\usepackage{jheppub}
\usepackage{latexsym,amsmath,amssymb,lmodern,float,url}
\usepackage{natbib}
\usepackage{color}
\usepackage{multirow}
\usepackage{bm}
\usepackage{array}

\def\de{\delta^{\vphantom{1}}}
\def\bde{{\bar\delta}}
\def\qq{{q\bar q}}
\def\QQ{{Q\bar Q}}

\def\bt{{\bar\theta}}
\def\hf{{\displaystyle{\frac 1 2}}}
\def\h3{{\displaystyle{\frac 3 2}}}
\newcommand{\Umat}{\mathbf{U}}
\newcommand{\Tmat}{\mathbf{T}}
\newcommand{\Fmat}{\mathbf{F}}
\newcommand{\Hmat}{\mathbf{H}}
\newcommand{\Rmat}{\mathbf{R}}

\newcommand{\Ns}{\mathcal{N}}
\newcommand{\diff}{\textrm{d}}

\begin{document}
\title{The Dynamical Diquark Model: First Numerical Results}
\author{Jesse F. Giron}
\emailAdd{jfgiron@asu.edu}
\author{Richard F. Lebed}
\emailAdd{Richard.Lebed@asu.edu}
\author{Curtis T. Peterson}
\emailAdd{curtistaylor.peterson@asu.edu}
\affiliation{Department of Physics, Arizona State University, Tempe,
AZ 85287, USA}

\date{March, 2019}

\abstract{
We produce the first numerical predictions of the dynamical diquark
model of multiquark exotic hadrons.  Using Born-Oppenheimer potentials
calculated numerically on the lattice, we solve coupled and uncoupled
systems of Schr\"{o}dinger equations to obtain mass eigenvalues for
multiplets of states that are, at this stage, degenerate in spin and
isospin.  Assuming reasonable values for these fine-structure
splittings, we obtain a series of bands of exotic states with a common
parity eigenvalue that agree well with the experimentally observed
charmoniumlike states, and we predict a number of other unobserved
states.  In particular, the most suitable fit to known pentaquark
states predicts states below the charmonium-plus-nucleon threshold.
Finally, we examine the strictest form of Born-Oppenheimer decay
selection rules for exotics and, finding them to fail badly, we
propose a resolution by relaxing the constraint that exotics must
occur as heavy-quark spin-symmetry eigenstates.
}

\maketitle

\section{Introduction}
\label{sec:Intro}

The existence of multiquark exotic hadrons is now a well-established
feature of strong-interaction physics.  At least 35 such states in the
heavy quark-antiquark ($Q \! = \! c$ or $b$) sector have been
experimentally established at various levels of statistical
significance.  Many have been observed beyond the $5\sigma$ level at
either multiple facilities, or through multiple production or decay
channels, or both.  A number of reviews in the recent
literature~\cite{Lebed:2016hpi,Chen:2016qju,Hosaka:2016pey,
Esposito:2016noz,Guo:2017jvc,Ali:2017jda,Olsen:2017bmm,
Karliner:2017qhf,Yuan:2018inv} summarize both experimental and
theoretical developments. 

Yet, even after many years of intensive study, no single theoretical
model has emerged to provide a successful, unified picture for
understanding the spectroscopy, decay patterns, and structure of these
novel states.  The most heavily studied alternatives in this regard,
including hadronic molecules, diquarks, hadroquarkonium, hybrids, and
kinematical threshold effects, both their benefits and drawbacks, are
amply discussed in the aforementioned reviews. The complete spectrum
might turn out to rely upon a delicate interplay of several of these
physical frameworks, meaning that each one must be fully understood
before a global model of the exotics can be confidently constructed.

In this work we employ a model in which the exotics are constructed
from quasi-bound heavy-light diquarks $\de , \bde$, which are formed
via the attractive channels ${\bf 3} \! \otimes \! {\bf 3} \! \to \!
\bar{\bf 3}$ [$\delta \! \equiv \! (Q q)_{\bar{\bf 3}}$] and $\bar{\bf
3} \! \otimes \! \bar{\bf 3} \! \to \! {\bf 3}$ [$\bde \! \equiv \!
(\bar Q {\bar q}^\prime)_{\bf 3}$] between the color-triplet quarks.
The most influential early application of diquarks to the problem of
heavy exotics~\cite{Maiani:2004vq} treats tetraquarks as bound ($\de
\bde$) molecules in a Hamiltonian formalism, using as interaction
operators the spin-spin couplings between the various component
quarks.  A later variant of this approach~\cite{Maiani:2014aja}
restricted the interactions to spin-spin couplings between quarks
within either the $\de$ or the $\bde$.
Reference~\cite{Anselmino:1992vg} provides a detailed review of
diquark phenomenology prior to the discovery of the heavy exotics.

Such an approach inspired the development of the {\it dynamical
diquark picture\/}~\cite{Brodsky:2014xia}, in which some of the light
quarks created in the production process of the $\QQ$ pair coalesce
with these heavy quarks to form a $\de$-$\bde$ pair.  Due to the large
energies available in either $b \! \to \! c$ or collider processes in
which exotics are produced, the $\de$-$\bde$ pair can achieve through
recoil a large spatial separation ($> \! 1$~fm) before being forced by
confinement either to form a single tetraquark state, or if the energy
is sufficiently high, for the color flux tube between the $\de$-$\bde$
pair to fragment to create a baryon-antibaryon pair.  The key feature
of this picture is a mechanism for producing multiquark states that
are spatially large yet strongly bound.  Indeed, the successive
accretion of additional quarks through the color-triplet binding
mechanism~\cite{Brodsky:2015wza} can be used to interpret pentaquark
states as {\it triquark}-diquark states, $\bt \de \! \equiv \! [\bar
Q(q_1 q_2)_{\bar{\bf 3}}]^{\vphantom{(}}_{\bf 3} (Q q_3)_{\bar{\bf
3}}$~\cite{Lebed:2015tna}.  The effectiveness of this mechanism is
clearly limited by competition from the attractive $\bar Q Q$ and
$\bar Q q$ channels, and a full theory of multiquark hadrons would
allow for including all possible configurations simultaneously.  One
important step in this much larger project is to uncover the
predictions of the $\de$-$\bde$ mechanism and see if the current data
provides support for the existence of such states.

The means by which the dynamical diquark {\em picture\/} may be
realized as a fully predictive {\em model\/}, including spectroscopy
and decay selection rules, is the subject of
Ref.~\cite{Lebed:2017min}.  In the original proposal of the
picture~\cite{Brodsky:2014xia}, the estimated size of the
$Z_c^-(4430)$ resonance appearing in $B^0 \! \to \! (\psi(2S) \pi^- )
K^+$ follows from taking a $\de$-$\bde$ pair (of known masses)
produced in the decay to recoil against the $K^+$.  Since the
diquarks, like quarks, are color triplets, a Cornell
Coulomb-plus-linear potential~\cite{Eichten:1978tg,Eichten:1979ms}
was assumed.  The final separation of the $\de$-$\bde$ pair upon
coming relatively to rest was calculated to be 1.16~fm.

The significance of this classical turning point in forming the
exotic state from the $\de$-$\bde$ pair was tied in
Refs.~\cite{Brodsky:2014xia,Lebed:2017min} to the
Wentzel-Kramers-Brillouin (WKB) approximation.  The WKB transition
wave function scales as $1/\sqrt{p}$, where $p$ is the classical
relative momentum of the constituents.  One therefore expects the
color flux tube between the $\de$-$\bde$ pair to stretch nearly to
its classical limit.  Such a state is spatially large but still
exhibits unscreened strong interactions between all of its
components.  It also possesses two heavy, slow-moving sources ($\de$
and $\bde$) connected by a lighter (mostly gluonic) field susceptible
to more rapid changes.  Analogous comments apply to $\bt$-$\de$
states.  These properties indicate that the system can be
characterized well by use of the Born-Oppenheimer (BO)
approximation~\cite{Born:1927boa}.

Although more familiar from its applications to atomic and molecular
systems, the BO approximation has also been implemented in particle
physics.  In fact, its use as a fundamental tool in lattice-QCD
calculations was initiated decades ago~\cite{Griffiths:1983ah}.  The
relevant physical observables are the energies of the light degrees of
freedom (d.o.f.) that connect a heavy, and hence static, $\QQ$ pair;
such energies as a function of $\QQ$ separation and orientation are
called {\it BO potentials}.  While multiple aspects of this problem in
strong-interaction physics have been studied in the intervening years,
the ones most relevant to the present work involve the calculation of
the BO potentials and its eigenvalues, which in turn give the masses
of heavy-quark hybrid mesons; short overviews of the key lattice
papers in this regard appear in
Refs.~\cite{Berwein:2015vca,Lebed:2017xih}.  For many years, the most
accurate lattice results of hybrid static potentials for substantial
$\QQ$ separation have been those of Juge, Kuti, and
Morningstar~\cite{Juge:1997nc,Juge:1999ie,Juge:2002br,
Morningstar:2019}.  Very recently, however, a new
collaboration~\cite{Capitani:2018rox} has begun to improve upon these
results.  In addition, high-quality simulations focusing upon small
$\QQ$ separations have been performed~\cite{Bali:2003jq}.  Also of
note are lattice simulations of two heavy quark-antiquark pairs, which
study the crossover between hadron molecule and $\de$-$\bde$
configurations~\cite{Bicudo:2017usw}.

A prototype of the approach underlying the present work is provided
by Ref.~\cite{Berwein:2015vca}, which supposes the known (neutral)
exotics to be hybrid mesons, and then develops an effective
field-theory formalism for computing their spectrum, using the BO
potentials as input to the Schr\"{o}dinger equations.  The first
treatment of multiquark exotic hadrons using the BO formalism
appeared in Refs.~\cite{Braaten:2013boa,Braaten:2014ita,
Braaten:2014qka}.  In these works, the valence light $\qq$ pair (for
tetraquarks) is treated as belonging to the light d.o.f.\@  In
contrast, the light quarks in the dynamical diquark model belong to
the diquarks, which in turn are treated as the heavy, pointlike
sources, while the light d.o.f.\ are purely gluonic (or include sea
quarks).

This paper carries out one of the central proposals of
Ref.~\cite{Lebed:2017min}: a numerical calculation of the spectrum of
$\de$-$\bde$ and $\bt$-$\de$ hidden-charm states, under the assumption
that their basic structure---at least in the last moments of their
evolution prior to decay---consists of heavy, slow-moving compact
diquarks interacting through the same hybrid BO potentials as those
appearing in the $\QQ$ sector.  We solve the resulting Schr\"{o}dinger
equations numerically and identify known exotic states with the
eigenstates of the lowest-lying BO potentials.  One can already
identify a number of assumptions implicit in this strategy; these and
several others of equal significance are discussed below.
Nevertheless, the initial results are quite encouraging: Choosing to
fix to either $X(3872)$ or $Z_c(4430)$ as a reference $\de$-$\bde$
state, one obtains a spectrum broadly consistent with the pattern of
the known tetraquark states, and for which the excited states above
the ground-state band [$\Sigma^+_g(1S)$] naturally have substantial
spatial extent.  In the pentaquark case, fixing to, {\it e.g.},
$P_c(4312)$ and $P_c(4457)$ predicts the masses of numerous unseen
states.

In carrying out calculations in this scheme, one must keep in mind the
central difficulty with diquark models (as discussed in any of the
reviews~\cite{Lebed:2016hpi,Chen:2016qju,Hosaka:2016pey,
Esposito:2016noz,Guo:2017jvc,Ali:2017jda,Olsen:2017bmm,
Karliner:2017qhf,Yuan:2018inv}): the proliferation of many more
potential multiquark states than have been observed to date.  For
example, the $J^{PC} \! = \! 1^{++}$ $X(3872)$ appears to lack an
isospin partner~\cite{Aubert:2004zr} that would arise from replacing
its $u \! \to \! d$ quarks.  On the other hand, the isotriplet
$1^{+-}$ $Z_c(3900)$ lies so close in mass to $X(3872)$ as to suggest
some sort of multiplet structure.  A truly complete diquark model must
explain both the absence of the former and the presence of the latter.
While we indicate how the details of this fine structure might be
resolved in Sec.~\ref{sec:Approx}, for the present work we seek only
to establish the broader pattern of multiplets, in particular, as
collections of states sorted in mass by their parity eigenvalue.

The organization of this paper is as follows: In
Sec.~\ref{sec:Spectrum} we reprise the notation for dynamical
diquark-model states developed in Ref.~\cite{Lebed:2017min}; the
reader unfamiliar with BO notation appropriate to the ``diatomic''
system is referred to Appendix~\ref{sec:BOapp}.
Section~\ref{sec:Schr} discusses the relevant Schr\"{o}dinger
equations, both uncoupled and coupled versions, as introduced in
Ref.~\cite{Berwein:2015vca}.  Details of our numerical approach to
solving these equations appear in Appendix~\ref{sec:Comp}.  In
Sec.~\ref{sec:Results} we present our results, and outline the
approximations used to obtain them in Sec.~\ref{sec:Approx},
describing how these simplifications can be lifted one by one.
Finally, Sec.~\ref{sec:Concl} presents our conclusions and directions
for subsequent development of the model.

\section{Spectrum of the Dynamical Diquark Model}
\label{sec:Spectrum}

\subsection{Tetraquarks}

The notation adopted in Ref.~\cite{Lebed:2017min} for states in the
dynamical diquark model begins with the notation introduced in
Ref.~\cite{Maiani:2014aja} for diquark-antidiquark ($\de$-$\bde$)
states of good total $J^{PC}$ with zero orbital angular momentum:
\begin{eqnarray}
J^{PC} = 0^{++}: & \; & X_0 \equiv \left| 0_\de , 0_\bde \right>_0 \,
, \ \
X_0^\prime \equiv \left| 1_\de , 1_\bde \right>_0 \, , \nonumber \\
J^{PC} = 1^{++}: & & X_1 \equiv \frac{1}{\sqrt 2} \left( \left| 1_\de
, 0_\bde \right>_1 \! + \left| 0_\de , 1_\bde \right>_1 \right) \, ,
\nonumber \\
J^{PC} = 1^{+-}: & & \  \, Z \equiv \frac{1}{\sqrt 2} \left( \left|
1_\de , 0_\bde \right>_1 \! - \left| 0_\de , 1_\bde \right>_1 \right)
\, , \nonumber \\
& & \; Z^\prime \equiv \left| 1_\de , 1_\bde \right>_1 \, ,
\nonumber \\
J^{PC} = 2^{++}: & & X_2 \equiv \left| 1_\de , 1_\bde \right>_2 \, .
\label{eq:Swavediquark}
\end{eqnarray}
The number before each $\de$($\bde$) subscript is the diquark
(antidiquark) spin, and the outer subscript on each ket is the total
quark spin $J$.  By straightforward use of $9j$ symbols, these states
can also be expressed in terms of states of good heavy-quark ($\QQ$)
and light-quark ($\qq$) spins [from which eigenvalues of the
charge-conjugation parity $C$ given in Eq.~(\ref{eq:Swavediquark})
are immediately determined].  The states in
Eq.~(\ref{eq:Swavediquark}) represent the tetraquark analogues to
heavy $S$-wave quark-model states such as $\eta_Q$ and $\psi \;
(\Upsilon)$.

One then allows for nonzero relative orbital angular momentum $L$
between the $\de$-$\bde$ pair.  Using the generic symbol $Y$ for
$X_0,X_1,Z,Z^\prime,X_2$ in Eq.~(\ref{eq:Swavediquark}), and ${\bf J}
\! = \! {\bf L} \! + \! {\bf S}$, where now $S$ is the total quark
spin, one obtains the states $Y^{(J)}_L$.  In terms of $C_Y$, the
$C$-parity of the underlying $S$-wave state $Y$, these states have
$P \! = \!  (-1)^L$ and $C \! = \! (-1)^L C_Y$.  Such states include
the analogues of the $P$-wave quark-model states $\chi_Q$, $h_Q$, as
well as the $D$-wave, $F$-wave, {\it etc.} states.  All of these
states also possess radial excitations, labeled by the quantum number
$n$.

Lastly, one appends the quantum numbers from the Born-Oppenheimer
(BO) excitation of the gluon field with respect to the quark state.
Given the BO potential with quantum numbers $\Gamma \! \equiv \!
\Lambda^\epsilon_\eta$ as defined in Appendix~\ref{sec:BOapp}, states
receive multiplicative factors to their $P$ and $C$ quantum numbers
of $\rho \! \equiv \! \epsilon (-1)^\Lambda$ and $\kappa \! \equiv \!
\eta \epsilon (-1)^\Lambda \! = \! \eta \rho$, respectively.  In
total, and suppressing the radial quantum number $n$, the physical
tetraquark eigenstates may be labeled $Y^{(J)\rho\kappa}_L$, where
\begin{equation}
P = \rho \, (-1)^L \, , \ \ C = \kappa \, (-1)^L C_Y \, .
\end{equation}
The resulting states associated with the lowest BO potentials (as
calculated on the lattice) are listed in Table~\ref{table:States}.  If
the light d.o.f.\ carry nonzero isospin $I$ [{\it e.g.}, $(cu)(\bar c
\bar d)$], then the $C$-parity eigenvalue of the state is replaced by
the $G$-parity eigenvalue, $G \! \equiv \! C (-1)^I$, where the $C$
eigenvalue is that of the neutral member of the isospin multiplet.
\begin{table*}
  \caption{Quantum numbers of the lowest tetraquark states expected in
  the dynamical diquark picture.  For each of the expected lowest
  Born-Oppenheimer potentials, the full multiplet for given $nL$
  eigenvalues is presented, using both the state notation developed in
  Ref.~\cite{Lebed:2017min} and the corresponding $J^{PC}$
  eigenvalues.  States with $J^{PC}$ not allowed for conventional
  $\qq$ mesons are indicated in boldface.}
\label{table:States}
\setlength{\extrarowheight}{1.5ex}
\begin{center}
\begin{tabular}{cccc}
  \hline\hline
  BO states & \multicolumn{3}{c}{State notation} \\
  \cline{2-4}
  & \multicolumn{3}{c}{State $J^{PC}$} \\
  \hline
  $\Sigma^+_g(1S)$ & $ X_{0 \, S}^{(0)++}$ &
  $Z_S^{(1) ++}$, $Z_S^{\prime \, (1) ++}$ &
  $X_{0 \, S}^{\prime \, (0) ++}$, $X_{1 \, S}^{(1) ++}$,
  $X_{2 \, S}^{(2) ++}$ \\
  & $0^{++}$ & $2 \times 1^{+-}$ & $[0,1,2]^{++}$ \\
  $\Sigma^+_g(1P)$ & $X_{0 \, P}^{(1) ++}$ &
  $[Z_P^{(0), \bm{(1)}, (2)}]^{++}$,
  $[Z_P^{\prime \, (0), \bm{(1)}, (2)}]^{++}$ &
  $X_{0 \, P}^{\prime \, (1) ++}$, \
  $[X_{1 \, P}^{\bm{(0)}, (1), (2)}]^{++}$, \
  $[X_{2 \, P}^{(1), (2), (3)}]^{++}$ \\
  & $1^{--}$ & $2 \times (0,\bm{1},2)^{-+}$ &
  $[ 1, \ (\bm{0},1,2), \ (1,2,3) ]^{--}$ \\
  $\Sigma^+_g(1D)$ & $X_{0 \, D}^{(2)++}$ &
  $[Z_D^{(1),\bm{(2)},(3)}]^{++}$,
  $[Z_D^{\prime \, (1),\bm{(2)},(3)}]^{++}$ &
  $X_{0 \, D}^{\prime \,  (2)++}$, \ 
  $[X_{1 \, D}^{(1),(2),(3)}]^{++}$, \
  $[X_{2 \, D}^{(0),(1),(2),(3),(4)}]^{++}$ \\
  & $2^{++}$ & $2 \times (1,\bm{2},3)^{+-}$
  & $[ 2, \ (1,2,3), \ (0,1,2,3,4)]^{++}$ \\
  $\Pi^+_u(1P)$ \& & $X_{0 \, P}^{(1) -+}$ &
  $[Z_P^{(0), (1), (2)}]^{-+}$,
  $[Z_P^{\prime \, (0), (1), (2)}]^{-+}$ &
  $X_{0 \, P}^{\prime \, (1) -+}$, \
  $[X_{1 \, P}^{\bm{(0)}, (1), \bm{(2)}}]^{-+}$, \
  $[X_{2 \, P}^{(1), \bm{(2)}, (3)}]^{-+}$ \\
  $\Sigma^-_u(1P)$ & $1^{+-}$ & $2 \times (0,1,2)^{++}$ &
  $[ 1, \ (\bm{0},1,\bm{2}), \ (1,\bm{2},3) ]^{+-}$ \\
  $\Pi^-_u(1P)$ & $X_{0 \, P}^{\bm{(1)} +-}$ &
  $[Z_P^{\bm{(0)}, (1), (2)}]^{+-}$,
  $[Z_P^{\prime \, \bm{(0)}, (1), (2)}]^{+-}$ &
  $X_{0 \, P}^{\prime \, \bm{(1)} +-}$, \
  $[X_{1 \, P}^{(0), \bm{(1)}, (2)}]^{+-}$, \
  $[X_{2 \, P}^{\bm{(1)}, (2), \bm{(3)}}]^{+-}$ \\
  & $\bm{1}^{-+}$ & $2 \times (\bm{0},1,2)^{--}$ &
  $[ \bm{1}, \ (0,\bm{1},2), \ (\bm{1},2,\bm{3}) ]^{-+}$ \\
  $\Sigma^-_u(1S)$ & $ X_{0 \, S}^{(0)-+}$ &
  $Z_S^{(1) -+}$, $Z_S^{\prime \, (1) -+}$ &
  $X_{0 \, S}^{\prime \, (0) -+}$, $X_{1 \, S}^{\bm{(1)} -+}$,
  $X_{2 \, S}^{(2) -+}$ \\
  & $0^{-+}$ & $2 \times 1^{--}$ & $[0,\bm{1},2]^{-+}$ \\
  $\Pi^+_u(1D)$ \& & $X_{0 \, D}^{(2)-+}$ &
  $[Z_D^{(1),(2),(3)}]^{-+}$,
  $[Z_D^{\prime \, (1),(2),(3)}]^{-+}$ &
  $X_{0 \, D}^{\prime \, (2)-+}$, \ 
  $[X_{1 \, D}^{\bm{(1)},(2),\bm{(3)}}]^{-+}$, \
  $[X_{2 \, D}^{(0),\bm{(1)},(2),\bm{(3)},(4)}]^{-+}$ \\
  $\Sigma^-_u(1D)$ & $2^{-+}$ & $2 \times (1,2,3)^{--}$
  & $[ 2, \ (\bm{1},2,\bm{3}), \ (0,\bm{1},2,\bm{3},4)]^{-+}$ \\
\hline\hline
\end{tabular}
\end{center}
\end{table*}

\subsection{Pentaquarks}

Much of the same construction holds for the $\bt$-$\de$ pentaquarks.
In that case, the states analogous to those in
Eq.~(\ref{eq:Swavediquark}) are denoted by~\cite{Lebed:2017min}:
\begin{eqnarray}
J^{P} = {\frac 1 2}^- \! \! : \ & &
P_{\frac 1 2} \equiv \left| \hf_\bt , 0_\de \right>_{\frac 1 2} , \ \
P^\prime_{\frac 1 2} \equiv \left| \hf_\bt , 1_\de \right>_{\frac 1 2}
, \nonumber \\
J^{P} = {\frac 3 2}^- \! \! : \ & &
P_{\frac 3 2} \equiv \left| \hf_\bt , 1_\de \right>_{\frac 3 2} .
\label{eq:SwaveDiTri}
\end{eqnarray}
The number before each $\bt$($\de$) subscript is the triquark
(diquark) spin, and the outer subscript on each ket is the total quark
spin $J$.  In this list, the light diquark internal to $\bt$ is
restricted to carry spin 0 (as well $ud$ flavor content with isospin
0), since all the known heavy pentaquark
candidates~\cite{Aaij:2019vzc}, $P_c(4312)$, $P_c(4380)$, $P_c(4440)$,
and $P_c(4457)$, appear in the decay of $\Lambda_b$, whose light $ud$
valence quarks carry these attributes.  In general, 6 additional
states, for which the light diquark in $\bt$ carries spin 1, can be
defined.

The relative orbital angular momentum $L$ and the BO potential
quantum numbers $\Gamma \! \equiv \! \Lambda^\epsilon$ are then
incorporated.  Since the $\bt$ and $\de$ components cannot form a
charge-conjugate pair, the core states of Eq.~(\ref{eq:SwaveDiTri})
are not $C$ eigenstates, and the BO potentials $\Lambda^\epsilon$
are not $\eta$ eigenstates (see Appendix~\ref{sec:BOapp}).  Defining
$\rho \! \equiv \! \epsilon (-1)^\Lambda$ as before, one obtains
\begin{equation}
P = \rho \, (-1)^{L+1} \, .
\end{equation}
Suppressing the radial quantum number $n$, the physical pentaquark
eigenstates may be labeled $P^{(J)\rho}_{SL}$, where $S$ now denotes
the total quark spin.  The resulting states associated with the
lowest BO potentials (as calculated on the lattice) are listed in
Table~\ref{table:States2}.
\begin{table*}
  \caption{Quantum numbers of the lowest pentaquark states expected in
  the dynamical triquark-diquark picture.  For each of the expected
  lowest Born-Oppenheimer potentials, the full multiplet for given
  $nL$ eigenvalues is presented, using both the state notation
  developed in Ref.~\cite{Lebed:2017min} and the corresponding $J^P$
  eigenvalues.}
\label{table:States2}
\setlength{\extrarowheight}{1.5ex}
\begin{center}
\begin{tabular}{ccc}
  \hline\hline
  BO states & \multicolumn{2}{c}{State notation} \\
  \cline{2-3}
  & \multicolumn{2}{c}{State $J^P$} \\
  \hline
  $\Sigma^+ (1S)$ & $P_{\frac 1 2 \, S}^{( \frac 1 2 )+}$,
  $P_{\frac 1 2 \, S}^{\prime \, ( \frac 1 2 )+}$ &
  $P_{\frac 3 2 \, S}^{( \frac 3 2 ) +}$ \\
  & $2 \times {\frac 1 2}^-_{\vphantom\dagger}$ & ${\frac 3 2}^-$ \\ 
  $\Sigma^+ (1P)$ & $\Big[ P_{\frac 1 2 \, P}^{( \frac 1 2 ),
  (\frac 3 2 )} \Big]^+, \ \Big[ P_{\frac 1 2 \, P}^{\prime
  \, ( \frac 1 2 ), (\frac 3 2)} \Big]^+$ &
  $\Big[ P_{\frac 3 2 \, P}^{( \frac 1 2 ), ( \frac 3 2 ),
  ( \frac 5 2 )} \Big]^+$ \\
  & $2 \times \left( \frac 1 2 , \frac 3 2 \right)^+
  _{\vphantom\dagger}$
  & $\left( \frac 1 2 , \frac 3 2 , \frac 5 2 \right)^+$ \\
  $\Sigma^+ (1D)$ & $\Big[ P_{\frac 1 2 \, D}^{( \frac 3 2 ),
  (\frac 5 2 )} \Big]^+, \ \Big[ P_{\frac 1 2 \, D}^{\prime
  \, ( \frac 3 2 ), (\frac 5 2)} \Big]^+$ &
  $\Big[ P_{\frac 3 2 \, D}^{( \frac 1 2 ), ( \frac 3 2 ),
  ( \frac 5 2 ), ( \frac 7 2 )} \Big]^+$ \\
  & $2 \times \left( \frac 3 2 , \frac 5 2 \right)^-
  _{\vphantom\dagger}$
  & $\left( \frac 1 2 , \frac 3 2 , \frac 5 2 , \frac 7 2 \right)^-$
  \\
  $\Pi^+ (1P)$ \& \ & $\Big[ P_{\frac 1 2 \, P}^{( \frac 1 2 ),
  (\frac 3 2 )} \Big]^-, \ \Big[ P_{\frac 1 2 \, P}^{\prime
  \, ( \frac 1 2 ), (\frac 3 2)} \Big]^-$ &
  $\Big[ P_{\frac 3 2 \, P}^{( \frac 1 2 ), ( \frac 3 2 ),
  ( \frac 5 2 )} \Big]^-$ \\
  $\Sigma^- (1P)$ & $2 \times \left( \frac 1 2 ,
  \frac 3 2 \right)^-_{\vphantom\dagger}$
  & $\left( \frac 1 2 , \frac 3 2 , \frac 5 2 \right)^-$ \\
  $\Pi^- (1P)$ & \multicolumn{2}{c}{Same as
  $\Sigma^+ (1P)_{\vphantom{\big[ }}$} \\
  $\Sigma^- (1S)$ & $P_{\frac 1 2 \, S}^{( \frac 1 2 )-}$,
  $P_{\frac 1 2 \, S}^{\prime \, ( \frac 1 2 )-}$ &
  $P_{\frac 3 2 \, S}^{( \frac 3 2 ) -}$ \\
  & $2 \times {\frac 1 2}^+_{\vphantom\dagger}$ & ${\frac 3 2}^+$ \\
  $\Pi^+ (1D)$  \& \ & $\Big[ P_{\frac 1 2 \, D}^{( \frac 3 2 ),
  (\frac 5 2 )} \Big]^-, \ \Big[ P_{\frac 1 2 \, D}^{\prime
  \, ( \frac 3 2 ), (\frac 5 2)} \Big]^-$ &
  $\Big[ P_{\frac 3 2 \, D}^{( \frac 1 2 ), ( \frac 3 2 ),
  ( \frac 5 2 ), ( \frac 7 2 )} \Big]^-$ \\
  $\Sigma^- (1D)$ & $2 \times \left( \frac 3 2 , \frac 5 2 \right)^+
  _{\vphantom\dagger}$
  & $\left( \frac 1 2 , \frac 3 2 , \frac 5 2 , \frac 7 2 \right)^+$
  \\
\hline\hline
\end{tabular}
\end{center}
\end{table*}

\section{Schr\"{o}dinger Equations for the Born-Oppenheimer
Potentials}
\label{sec:Schr}

At its core, the calculation of the spectrum of hybrids in
Ref.~\cite{Berwein:2015vca} amounts to the use of the BO potentials
calculated on the lattice in Refs.~\cite{Juge:2002br,Bali:2003jq} to
find the energy eigenvalues of Schr\"{o}dinger equations between a
static $\QQ$ pair ($c\bar c$, $c\bar b$, and $b\bar b$ are all
considered).  The relevant Schr\"{o}dinger equations actually arise
directly from QCD through a systematic $1/m_Q$ expansion by the
application of effective field theories: first
NRQCD~\cite{Caswell:1985ui,Bodwin:1994jh} (in which the hard scale
$m_Q$ is integrated out), and then pNRQCD~\cite{Pineda:1997bj,
Brambilla:1999xf} (in which the softer scale of momentum transfer
between the $\QQ$ pair is integrated out).  The gluonic pNRQCD static
energies between the $\QQ$ pair are then none other than the BO
potentials, which are obtained numerically on the lattice.

Since the fundamental quark mass $m_Q$ appears directly in the
analysis of Ref.~\cite{Berwein:2015vca}, the authors take care to
identify the details of their renormalization scheme for both $m_Q$
and the perturbative short-distance behavior of the potential between
the fundamental $\QQ$ pair.  In our case, the corresponding mass
$m_\de$  is that of the diquark (or $m_\bt$ for the triquark), which
is of course not a fundamental Lagrangian parameter, and therefore in
this analysis $m_\de$, $m_\bt$, and the $\de$-$\bde$ and $\bt$-$\de$
potentials are treated purely phenomenologically.

More central to the current calculation is that the full
Schr\"{o}dinger equations for the ``diatomic'' system contain not
one, but two special points, corresponding to the two
heavy-constituent positions, separated by a distance $r$.  One
expects additional symmetries between the static energies to arise in
the limit $r \! \to \! 0$, where the cylindrical $D_{\infty h}$
symmetry for the ``homonuclear'' $\de$-$\bde$ case or the conical
$C_{\infty v}$ symmetry for the ``heteronuclear'' $\bt$-$\de$ case
(see Appendix~\ref{sec:BOapp} and Fig.~\ref{fig:BO_Quantum}) is
supplanted by the higher spherical $O(3)$ symmetry.  These $r \! \to
\! 0$ static energy configurations, transforming as color adjoints in
the light d.o.f.\ and corresponding to degenerate BO potentials in
the limit $r \! \to \! 0$, are called {\it gluelumps}.  One finds,
for instance, that the $\de$-$\bde$ $\Sigma^-_u$ and $\Pi^+_u$ BO
potentials approach a single gluelump with $J^{PC} \! = \! 1^{+-}$.
Additionally, the loss of the single (spherical) symmetry center for
$r \! > \! 0$ means that the angular part of the Laplacian in the
Schr\"{o}dinger equation is no longer solved by familiar spherical
harmonics, but by slightly more complicated forms~\cite{Landau:1977}.

All $\de$-$\bde$ BO potentials arising from a particular glue\-lump
appear together in a coupled system of Schr\"{o}dinger equations.  The
ground-state BO potential $\Sigma^+_g$ does not mix with others, and
therefore appears in an uncoupled equation.  While the degenerate BO
potentials $\Pi^+_u$ and $\Pi^-_u$ carry opposite $\epsilon$ parities,
only $\Pi^+_u(nL)$ produces states with the same $J^{PC}$ quantum
numbers as $\Sigma^-_u(nL)$ for $L \! > \! 0$ [$\Pi^+_u(nS)$ is
forbidden by Eq.~(\ref{eq:LambdaMax})].  Indeed, $\Pi^+_u$ is found in
lattice simulations to approach $\Sigma^-_u$ as $r \! \to \!  0$, to
form the $1^{+-}$ gluelump~\cite{Brambilla:1999xf}.  The
Schr\"{o}dinger equations for $\Pi^+_u(nL)$ and $\Sigma^-_u(nL)$ with
$L \! > \! 0$ therefore must be solved as a coupled system, while
those for $\Sigma^+_g(nL)$, $\Pi^-_u(nL)$ (with $L \! > \! 0$), or
$\Sigma^-_u(nS)$ remain uncoupled.  One finds different mass
eigenvalues emerging from $\Pi^\pm_u(1P)$, a lifting of the parity
symmetry called $\Lambda$-$\it{doubling}$~\cite{Landau:1977}.
Higher-mass gluelumps have been found to split into even more BO
potentials~\cite{Brambilla:1999xf}.  For example, the $2^{--}$
gluelump supports the BO potentials $\Sigma^-_g$, $\Pi^{+\prime}_g$,
and $\Delta^-_g$; its Schr\"{o}dinger equations for $D$-wave solutions
would couple all three of them.  Analogous comments apply to the
$\bt$-$\de$ BO potentials, once the $\eta \! = \! g,u$ subscript is
removed.

The radial Schr\"{o}dinger equations for the uncoupled BO potentials
$V_\Gamma$ assume the conventional form (with $\hbar \! = \! 1$):
\begin{equation} \label{eq:uncoupled}
\left[ -\frac{1}{2\mu r^2} \partial_r r^2 \partial_r  \! +
\frac{\ell(\ell + 1)}{2\mu r^2} + V_\Gamma (r) \right]
\! \psi_\Gamma^{(n)} (r) = E_n \psi_\Gamma^{(n)} (r) \, ,
\end{equation}
and for the coupled potentials $\Pi^+_u, \Sigma^-_u$, they read:
\begin{eqnarray}
\lefteqn{\left[ -\frac{1}{2\mu r^2} \partial_r r^2 \partial_r  \! +
\frac{1}{2\mu r^2} \left(
\begin{array}{cc} \ell (\ell + 1) + 2 & 2 \sqrt{\ell (\ell + 1)} \\
2 \sqrt{\ell (\ell + 1)} & \ell (\ell + 1) 
\end{array}
\right) \right. } \nonumber \\
& & + \left. \left(
\begin{array}{cc} V^{\vphantom\dagger}_{\Sigma^-_u} (r) & 0 \\ 0 &
V^{\vphantom\dagger}_{\Pi^+_u} (r)
\end{array} \right) \right] \!
\left( \begin{array}{c} \psi^{(n)}_{\Sigma^-_u} (r) \\
\psi^{(n)^{\vphantom\dagger}}_{\Pi^+_u} (r) \end{array} \right) \! =
E_n  \! \left( \begin{array}{c} \psi^{(n)}_{\Sigma^-_u} (r) \\
\psi^{(n)^{\vphantom\dagger}}_{\Pi^+_u} (r) \end{array} \right) .
\nonumber \\ & & 
\label{eq:coupled}
\end{eqnarray}
The details of the numerical methods for solving these coupled
Schr\"{o}dinger equations, using the state-of-the-art techniques of
Refs.~\cite{Johnson:1978} and~\cite{Hutson:1994}, are discussed in
Appendix~\ref{sec:Comp}.

\section{Results}
\label{sec:Results}

Using the techniques of Sec.~\ref{sec:Schr}, one predicts the full
spectrum of states in the dynamical diquark model with only two
further inputs: the diquark masses $m_\de$ and $m_\bde$ (or triquark
mass $m_\bt$) and specific functional forms $V(r)$ for the BO
potentials $\Gamma$.  The potentials are intrinsically
nonperturbative in nature and can only be computed from first
principles by using lattice QCD simulations.  In this regard we apply
the results of Refs.~\cite{Juge:1997nc,Juge:1999ie,Juge:2002br,
Morningstar:2019}, especially the online summary of results in
Ref.~\cite{Morningstar:2019}, to which we refer as JKM; and
separately, we apply the results of the very recent calculations of
Ref.~\cite{Capitani:2018rox}, to which we refer as CPRRW\@.
Furthermore, since the corresponding calculation for the ground-state
($\Sigma_g^+$) BO potential for the $c\bar c$ system with both
component masses given by $m_c$ would generate the conventional
charmonium spectrum, we also include for $\Sigma_g^+$ cases the
phenomenological Cornell potential used in the fit of
Ref.~\cite{Barnes:2005pb} (but suppressing spin-dependent couplings),
to which we refer as BGS.

The numerical results are summarized in Tables~\ref{table:BestFit},
\ref{table:Fit2}, and \ref{table:Fit3} for hidden-charm tetraquarks
and Table~\ref{table:FitPenta} for hidden-charm pentaquarks.  We take
$m_\bde \! = \! m_\de$ in each case.  Fine-structure spin-dependent
mass splittings among the states of each level $\Gamma(nL)$ (as
enumerated in Tables~\ref{table:States} and \ref{table:States2}) are
neglected in the present calculation, and the approach to include them
in future calculations is discussed in detail in
Sec.~\ref{sec:Approx}.

\begin{table*}[ht]
\caption{Mass eigenvalues $M$ (in GeV) for hidden-charm dynamical
diquark states that are eigenstates of the indicated BO potentials
corresponding to quantum numbers $nL$, for given diquark masses
$m_{\delta}$ (in GeV).  The particular form of the BO potential used
is that given by lattice simulations JKM~\cite{Morningstar:2019} or
CPRRW~\cite{Capitani:2018rox}, or (for the $\Sigma^+_g$ potential) by
the Cornell potential obtained from a fit BGS~\cite{Barnes:2005pb} to
conventional charmonium.  Also predicted are the corresponding
expectation values for the length scales ${\langle 1/r \rangle}^{-1}$
and $\langle r \rangle$ (in fm).  Fixing to the experimental mass of
$X(3872)$ or $Z_c^-(4430)$ predicts $m_{\delta}$ and the whole
spectrum, either under the assumption that $X(3872)$ is a
$\Sigma^+_g(1S)$ state or that $Z_c(4430)$ is a $\Sigma^+_g(2S)$ state
(as indicated by boldface).}
\label{table:BestFit}
\setlength{\extrarowheight}{0.075ex}
\begin{tabular}{c c c c c c @{\hskip 0.5em} | c c c c }
\hline\hline
\multicolumn{2}{c}{\multirow{2}{*}{}}&\multicolumn{4}{c|}{$X(3872)$}&
\multicolumn{4}{c}{$Z_c(4430)$}\\\cline{3-10}\hline
BO states & Potential & $M$ & $m_{\delta}$&
${\langle}{1/r}{\rangle}^{-1}$&${\langle}{r}{\rangle}$ &$M$&
$m_{\delta}$&${\langle}{1/r}{\rangle}^{-1}$&
${\langle}{r}{\rangle}$\\
\hline
\multicolumn{1}{ c }{{$\Sigma_g^+(1S)$}}
&JKM& $\mathbf{3.8711}$ & $1.8747$ & $0.27202$ & $0.36485$ & $3.9077$
& $1.8946$ & $0.27075$ & $0.36322$ \\
&CPRRW& $\mathbf{3.8721}$ & $1.8535$ & $0.27519$ & $0.36904$
& $3.9108$ & $1.8745$ & $0.27384$ & $0.36694$ \\
&BGS&$\mathbf{3.8718}$&$1.9402$& $0.21347$ &$0.30268$ & $3.8824$&
$1.9462$& $0.21301$ &$0.30221$ \\
\multicolumn{1}{ c }{{$\Sigma_g^+(2S)$}}
&JKM& $4.4430$ & $1.8747$ & $0.42698$ & $0.69081$ & $\mathbf{4.4782}$
& $1.8946$ & $0.42524$ & $0.68825$ \\
&CPRRW& $4.4410$ & $1.8535$ & $0.43057$ & $0.69640$ &
$\mathbf{4.4781}$ & $1.8745$ & $0.42877$ & $0.69360$ \\
&BGS&$4.4674$&$1.9402$& $0.42621$ &$0.69756$ &$\mathbf{4.4781} $&
$1.9462$& $0.42562$ &$0.69640$ \\
\multicolumn{1}{ c }{{$\Sigma_g^+(1P)$}}&JKM& $4.2457$
& $1.8747$ & $0.48968$ & $0.56601$ & $4.2816$ & $1.8946$ & $0.48773$ &
$0.56392$ \\
&CPRRW& $4.2435$ & $1.8535$ & $0.49379$ & $0.57067$ & $4.2814$ &
$1.8745$ & $0.49170$ & $0.56834$ \\
&BGS& $4.3471$ & $1.9402$ & $0.48361$ & $0.56787$ & $4.3580$ &
$1.9462$ & $0.48285$ & $0.56718$ \\
\multicolumn{1}{ c }{{$\Sigma_g^+(2P)$}}
&JKM& $4.7128$ & $1.8747$ & $0.62445$ & $0.84285$ & $4.7473$ &
$1.8946$ & $0.62201$ & $0.83982$ \\
&CPRRW& $4.7092$ & $1.8535$ & $0.62911$ & $0.84913$ & $4.7456$ &
$1.8745$ & $0.62664$ & $0.84564$ \\
&BGS& $4.7416$ & $1.9402$ & $0.65333$ & $0.89663$ & $4.7523$ &
$1.9462$ & $0.65243$ & $0.89547$ \\
\multicolumn{1}{ c }{{$\Sigma_g^+(1D)$}}
&JKM& $4.5318$ & $1.8747$ & $0.66414$ & $0.73132$ & $4.5669$ &
$1.8946$ & $0.66168$ & $0.72853$ \\
&CPRRW& $4.5282$ & $1.8535$ & $0.66921$ & $0.73668$ & $4.5653$ &
$1.8745$ & $0.66651$ & $0.73365$ \\
&BGS& $4.6151$ & $1.9402$ & $0.69780$ & $0.77323$ & $4.6259$ &
$1.9462$ & $0.69690$ & $0.77230$ \\
\multicolumn{1}{ c }{{$\Sigma_g^+(2D)$}}
&JKM& $4.9476$ & $1.8747$ & $0.78634$ & $0.98022$ & $4.9813$ &
$1.8946$ & $0.78332$ & $0.97672$ \\
&CPRRW& $4.9431$ & $1.8535$ & $0.79199$ & $0.98697$ & $4.9787$ &
$1.8745$ & $0.78879$ & $0.98348$ \\
&BGS& $4.9486$ & $1.9402$ & $0.84597$ & $1.0645$ & $4.9592$ &
$1.9462$ & $0.84497$ & $1.0633$ \\
\multicolumn{1}{ c }{$\Pi_u^+(1P)$ \& }
&JKM& $4.9156$ & $1.8747$ & $0.44931$ & $0.56950$ & $4.9539$ &
$1.8946$ & $0.44833$ & $0.56834$ \\
\ \ $\Sigma_u^- (1P)$ &CPRRW& $4.8786$ & $1.8535$ & $0.44614$ &
$0.56438$ & $4.9190$ & $1.8745$ & $0.44512$ & $0.56298$ \\
\multicolumn{1}{ c }{$\Pi_u^+(2P)$ \& }
&JKM& $5.2281$ & $1.8747$ & $0.54325$ & $0.84052$ & $5.2648$ &
$1.8946$ & $0.54181$ & $0.83819$ \\
\ \ $\Sigma_u^- (2P)$ &CPRRW& $5.2066$ & $1.8535$ & $0.52965$ &
$0.81887$ & $5.2450$ & $1.8745$ & $0.52816$ & $0.81677$ \\
\multicolumn{1}{ c }{{$\Pi_u^-(1P)$}}&JKM& $5.0291$ &
$1.8747$ & $0.66230$ & $0.74739$ & $5.0667$ & $1.8946$ & $0.66066$ &
$0.74552$ \\
&CPRRW& $4.9949$ & $1.8535$ & $0.65075$ & $0.73435$ & $5.0344$ &
$1.8745$ & $0.64908$ & $0.73225$ \\
\multicolumn{1}{ c }{{$\Pi_u^-(2P)$}}&JKM& $5.3701$ &
$1.8747 $ & $0.74501$ & $0.98068$ & $5.4060$ & $1.8946$ & $0.74307$ &
$0.97789$ \\
&CPRRW& $5.3564$ & $1.8535$ & $0.71810$ & $0.94716$ & $5.3939$ &
$1.8745$ & $0.71619$ & $0.94436$ \\
\multicolumn{1}{ c }{{$\Sigma_u^-(1S)$}}
&JKM& $5.3139$ & $1.8747 $ & $0.58948$ & $0.63819$ & $5.3507$ &
$1.8946$ & $0.58803$ & $0.63609$ \\
&CPRRW& $5.2897$ & $1.8535$ & $0.56550$ & $0.64448$ & $5.3285$ &
$1.8745$ & $0.56357$ & $0.64238$ \\
\multicolumn{1}{ c }{{$\Sigma_u^-(2S)$}}
&JKM& $5.7375$ & $1.8747 $ & $0.74128$ & $0.88755$ & $5.7725$ &
$1.8946$ & $0.73975$ & $0.88452$ \\
&CPRRW& $5.7105$ & $1.8535$ & $0.66424$ & $0.88336$ & $5.7473$ &
$1.8745$ & $0.66209$ & $0.88080$ \\
\multicolumn{1}{ c }{$\Pi_u^+(1D)$ \&}
&JKM& $5.1028$ & $1.8747$ & $0.66444$ & $0.75321$ & $5.1401$ &
$1.8946$ & $0.66280$ & $0.75134$ \\
\ \ $\Sigma_u^- (1D)$ &CPRRW& $5.0718$ & $1.8535$ & $0.65632$ &
$0.74296$ & $5.1110$ & $1.8745$ & $0.65472$ & $0.74110$ \\
\multicolumn{1}{ c }{$\Pi_u^+(2D)$ \&}
&JKM& $5.4114$ & $1.8747$ & $0.74038$ & $0.97253$ & $5.4471$ &
$1.8946$ & $0.73835$ & $0.96974$ \\
\ \  $\Sigma_u^- (2D)$ &CPRRW& $5.4012$ & $1.8535$ & $0.71834$ &
$0.94157$ & $5.4386$ & $1.8745$ & $0.71631$ & $0.93901$ \\
\hline
\end{tabular}
\end{table*}

\begin{table*}[ht]
\caption{As in Table~\ref{table:BestFit}, except now assuming that
$X(3872)$ is a $\Sigma^+_g(1D)$ state or that $Z_c(4430)$ is a
$\Sigma^+_g(2D)$ state (as indicated by boldface) in order to fix the
full spectrum.}
\label{table:Fit2}
\setlength{\extrarowheight}{0.46ex}
\begin{tabular}{c c c c c c @{\hskip 0.5em} | c c c c }
\hline\hline
\multicolumn{2}{c}{\multirow{2}{*}{}}&\multicolumn{4}{c}{$X(3872)$}&
\multicolumn{4}{c}{$Z_c(4430)$}\\\cline{3-10}\hline
BO states & Potential & $M$ & $m_{\delta}$&
${\langle}{1/r}{\rangle}^{-1}$&${\langle}{r}{\rangle}$&$M$&
$m_{\delta}$&${\langle}{1/r}{\rangle}^{-1}$&
${\langle}{r}{\rangle}$\\
\hline
\multicolumn{1}{ c }{{$\Sigma_g^+(1S)$}}
&JKM& $3.1759$ & $1.4925$ & $0.30043$ & $0.40140$ & $3.3552$ &
$1.5921$ & $0.29221$ & $0.39116$ \\
&CPRRW& $3.1809$ & $1.4734$ & $0.30392$ & $0.40559$ & $3.3604$ &
$1.5731$ & $0.29555$ & $0.39488$ \\
&BGS& $3.1373$ & $1.5208$ & $0.25111$ & $0.35251$ & $3.3948$ &
$1.6690$ & $0.23635$ & $0.33295$ \\
\multicolumn{1}{ c }{{$\Sigma_g^+(2S)$}}
&JKM& $3.7824$ & $1.4925$ & $0.46502$ & $0.75111$ & $ 3.9515$ &
$1.5921$ & $0.45401$ & $0.73342$ \\
&CPRRW& $3.7848$ & $1.4734$ & $0.46919$ & $0.75717$ & $3.9539$ &
$1.5731$ & $0.45789$ & $0.73947$ \\
&BGS& $3.7323$ & $1.5208$ & $0.47448$ & $0.77765$ & $3.9888$ &
$1.6690$ & $0.45565$ & $0.74646$ \\
\multicolumn{1}{ c }{{$\Sigma_g^+(1P)$}}
&JKM& $3.5670$ & $1.4925$ & $0.53301$ & $0.61560$ & $3.7414$ &
$1.5921$ & $0.52048$ & $0.60117$ \\
&CPRRW& $3.5692$ & $1.4734$ & $0.53754$ & $0.62073$ & $3.7436$ &
$1.5731$ & $0.52480$ & $0.60629$ \\
&BGS& $3.5953$ & $1.5208$ & $0.54134$ & $0.63377$ & $3.8582$ &
$1.6690$ & $0.51884$ & $0.60815$ \\
\multicolumn{1}{ c }{{$\Sigma_g^+(2P)$}}
&JKM& $4.0675$ & $1.4925$ & $0.67744$ & $0.91316$ & $4.2322$ &
$1.5921$ & $0.66199$ & $0.89267$ \\
&CPRRW& $4.0686$ & $1.4734$ & $0.68282$ & $0.92015$ & $4.2330$ &
$1.5731$ & $0.66713$ & $0.89919$ \\
&BGS& $4.0111$ & $1.5208$ & $0.72160$ & $0.98674$ & $4.2655$ &
$1.6690$ & $0.69475$ & $0.95181$ \\
\multicolumn{1}{ c }{{$\Sigma_g^+(1D)$}}
&JKM& $\mathbf{3.8714}$ & $1.4925$ & $0.72015$ & $0.79256$ & $4.0404$
& $1.5921$ & $0.70409$ & $0.77486$ \\
&CPRRW& $\mathbf{3.8724}$ & $1.4734$ & $0.72575$ & $0.79907$ &
$4.0413$ & $1.5731$ & $0.70921$ & $0.78091$ \\
&BGS&$\mathbf{3.8716}$&$1.5208$& $0.77026$ & $0.85216$ & $4.1311$ &
$1.6690$& $0.74205$ & $0.82189$\\
\multicolumn{1}{ c }{{$\Sigma_g^+(2D)$}}
&JKM& $4.3181$ & $1.4925$ & $0.85150$ & $1.0608$ \, & $\mathbf{4.4781}$
& $1.5921$ & $0.83268$ & $1.0370$ \, \\
&CPRRW& $4.3184$ & $1.4734$ & $0.85796$ & $1.0687$ \, &
$\mathbf{4.4782}$ & $1.5731$ & $0.83853$ & $1.0449$ \, \\
&BGS& $4.2276$ & $1.5208$ & $0.92844$ & $1.1664$ \, &
$\mathbf{4.4782}$ & $1.6690$& $0.89628$ & $1.1269$ \, \\
\multicolumn{1}{ c }{{$\Pi_u^+(1P)$ \& }}
&JKM& $4.1850$ & $1.4925$ & $0.47177$ & $0.59931$ & $4.3742$ &
$1.5921$ & $0.46523$ & $0.59046$ \\
$\Sigma_u^- (1P)$ &CPRRW& $4.1537$ & $1.4734$ & $0.46796$ & $0.59325$
& $4.3426$ & $1.5731$ & $0.46163$ & $0.58440$ \\
\multicolumn{1}{ c }{{$\Pi_u^+(2P)$ \& }}
&JKM& $4.5358$ & $1.4925$ & $0.57713$ & $0.89221$ & $4.7137$ &
$1.5921$ & $0.56736$ & $0.87731$ \\
$\Sigma_u^- (2P)$ &CPRRW& $4.5237$ & $1.4734$ & $0.56085$ & $0.86660$
& $4.7001$ & $1.5731$ & $0.55191$ & $0.85263$ \\
\multicolumn{1}{ c }{{$\Pi_u^-(1P)$}}&JKM& $4.3156$ &
$1.4925$ & $0.69973$ & $0.79023$ & $4.4998$ & $1.5921$ & $0.68874$ &
$0.77765$ \\
&CPRRW& $4.2886$ & $1.4734$ & $0.68719$ & $0.77579$ & $4.4719$ &
$1.5731$ & $0.67659$ & $0.76368$ \\
\multicolumn{1}{ c }{{$\Pi_u^-(2P)$}}&JKM& $4.6956$ &
$1.4925$ & $0.78923$ & $1.0408$ \, & $4.8683$ & $1.5921$ & $0.77667$ &
$1.0231$ \, \\
&CPRRW& $4.6945$ & $1.4734$ & $0.75910$ & $1.0016$ \, & $4.8647$ &
$1.5731$ & $0.74721$ & $0.98580$ \\
\multicolumn{1}{ c }{{$\Sigma_u^-(1S)$}}
&JKM& $4.6183$ & $1.4925$ & $0.62318$ & $0.68173$ & $4.7631$ &
$1.5921$ & $0.61497$ & $0.67148$ \\
&CPRRW& $4.5982$ & $1.4734$ & $0.60441$ & $0.68871$ & $4.7773$ &
$1.5731$ & $0.59306$ & $0.67567$ \\
\multicolumn{1}{ c }{{$\Sigma_u^-(2S)$}}
&JKM& $5.0816$ & $1.4925$ & $0.77807$ & $0.95134$ & $5.2169$ &
$1.5921$ & $0.76916$ & $0.93644$ \\
&CPRRW& $5.0626$ & $1.4734$ & $0.70757$ & $0.94110$ & $5.2288$ &
$1.5731$ & $.69475$ & $0.92434$ \\
\multicolumn{1}{ c }{{$\Pi_u^+(1D)$ \& }}
&JKM& $4.3956$ & $1.4925$ & $0.70133$ & $0.79582$ & $4.5780$ &
$1.5921$ & $0.69051$ & $0.78324$ \\
$\Sigma_u^- (1D)$ &CPRRW& $4.3715$ & $1.4734$ & $0.69184$ & $0.78324$
& $4.5531$ & $1.5731$ & $0.68131$ & $0.77160$ \\
\multicolumn{1}{ c }{{ $\Pi_u^+(2D)$ \& }}
&JKM& $4.7404$ & $1.4925$ & $0.78720$ & $1.0333$ \, & $4.9121$ &
$1.5921$ & $0.77359$ & $1.0156$ \, \\
$\Sigma_u^- (2D)$ &CPRRW& $4.7424$ & $1.4734$ & $0.76152$ & $0.99698$
& $4.9117$ & $1.5731$ & $0.74877$ & $0.98068$ \\
\hline
\end{tabular}
\end{table*}

\begin{table*}[ht]
\caption{As in Table~\ref{table:BestFit}, except now assuming that
$X(3872)$ is a $\Pi^+_u(1P)$-$\Sigma^-_u(1P)$ state or that
$Z_c(4430)$ is a $\Pi^+_u(2P)$-$\Sigma^-_u(2P)$ state (as indicated by
boldface) in order to fix the full spectrum.  Since the $\Sigma^+_g$
potential is not used here to fix to the mass of a physical state, no
BGS~\cite{Barnes:2005pb} fit is included.}
\label{table:Fit3}
\setlength{\extrarowheight}{1.0ex}
\begin{tabular}{c c c c c c @{\hskip 0.5em} | c c c c }
\hline\hline
\multicolumn{2}{c}{\multirow{2}{*}{}}&\multicolumn{4}{c}{$X(3872)$}&
\multicolumn{4}{c}{$Z_c(4430)$}\\\cline{3-10}
\hline
BO states & Potential & $M$ & $m_{\delta}$&
${\langle}{1/r}{\rangle}^{-1}$&${\langle}{r}{\rangle}$&$M$&
$m_{\delta}$&${\langle}{1/r}{\rangle}^{-1}$&
${\langle}{r}{\rangle}$\\
\hline
\multicolumn{1}{ c }{{$\Sigma_g^+(1S)$}}
&JKM& $2.8809$ & $1.3266$ & $0.31585$ & $0.42096$ & $3.1177$ &
$1.4600$ & $0.30327$ & $0.40513$ \\
&CPRRW& $2.9148$ & $1.3238$ & $0.31791$ & $0.42375$ & $3.1345$ &
$1.4475$ & $0.30621$ & $0.40838$ \\
\multicolumn{1}{ c }{{$\Sigma_g^+(2S)$}}
&JKM& $3.5068$ & $1.3266$ & $0.48586$ & $0.78324$ & $3.7277$ &
$1.4600$ & $0.46888$ & $0.75670$ \\
&CPRRW& $3.5364$ & $1.3238$ & $0.48818$ & $0.78697$ & $3.7413$ &
$1.4475$ & $0.47229$ & $0.76182$ \\
\multicolumn{1}{ c }{{$\Sigma_g^+(1P)$}}
&JKM& $3.2817$ & $1.3266$ & $0.55663$ & $0.64261$ & $3.5106$ &
$1.4600$ & $0.53727$ & $0.62026$ \\
&CPRRW& $3.3120$ & $1.3238$ & $0.55909$ & $0.64541$ & $3.5243$ &
$1.4475$ & $0.54106$ & $0.62492$ \\
\multicolumn{1}{ c }{{$\Sigma_g^+(2P)$}}
&JKM& $3.8005$ & $1.3266$ & $0.70618$ & $0.95181$ & $4.0145$ &
$1.4600$ & $0.68261$ & $0.92061$ \\
&CPRRW& $3.8280$ & $1.3238$ & $0.70897$ & $0.95554$ & $4.0264$ &
$1.4475$ & $0.68719$ & $0.92573$ \\
\multicolumn{1}{ c }{{$\Sigma_g^+(1D)$}}
&JKM& $3.5962$ & $1.3266$ & $0.75087$ & $0.82608$ & $3.8168$ &
$1.4600$ & $0.72575$ & $0.79907$ \\
&CPRRW& $3.6244$ & $1.3238$ & $0.75377$ & $0.82934$ & $3.8290$ &
$1.4475$ & $0.73044$ & $0.80373$ \\
\multicolumn{1}{ c }{{$\Sigma_g^+(2D)$}}
&JKM& $4.0599$ & $1.3266$ & $0.88702$ & $1.1050$ \, & $4.2667$ &
$1.4600$ & $0.85796$ & $1.0687$ \, \\
&CPRRW& $4.0858$ & $1.3238$ & $0.89033$ & $1.1087$ \, & $4.2775$ &
$1.4475$ & $0.86314$ & $1.0752$ \, \\
\multicolumn{1}{ c }{{$\Pi_u^+(1P)$ \& }}
&JKM& $\mathbf{3.8722}$ & $1.3266$ & $0.48410$ & $0.61514$ & $4.1234$
& $1.4600$ & $0.47395$ & $0.60210$ \\
$\Sigma_u^- (1P)$ &CPRRW& $\mathbf{3.8723}$ & $1.3238$ & $0.47881$ &
$0.60722$ & $4.1048$ & $1.4475$ & $0.46970$ & $0.59558$ \\
\multicolumn{1}{ c }{{$\Pi_u^+(2P)$ \& }}
&JKM& $4.2444$ & $1.3266$  & $0.59570$ & $0.92015$ &
$\mathbf{4.4782}$ & $1.4600$ & $0.58056$ & $0.89733$ \\
$\Sigma_u^- (2P)$ &CPRRW& $4.2637$ & $1.3238$ & $0.57635$ & $0.88988$
& $\mathbf{4.4782}$ & $1.4475$ & $0.56350$ & $0.87032$ \\
\multicolumn{1}{ c }{{$\Pi_u^-(1P)$}}
&JKM& $4.0125$ & $1.3266$ & $0.72015$ & $0.81398$ & $4.2559$ &
$1.4600$ & $0.70340$ & $0.79442$ \\
&CPRRW& $4.0168$ & $1.3238$ & $0.70479$ & $0.79628$ & $4.2412$ &
$1.4475$ & $0.69007$ & $0.77905$ \\
\multicolumn{1}{ c }{{$\Pi_u^-(2P)$}}
&JKM& $4.4142$ & $1.3266$ & $0.81375$ & $1.0729$ \, & $4.6398$ &
$1.4600$ & $0.79389$ & $1.0463$ \, \\
&CPRRW& $4.4453$ & $1.3238$ & $0.77892$ & $1.0286$ \, & $4.6508$ &
$1.4475$ & $0.76233$ & $1.0058$ \, \\
\multicolumn{1}{ c }{{$\Sigma_u^-(1S)$}}
&JKM& $4.3383$ & $1.3266$ & $0.62300$ & $0.71013$ & $4.5743$ &
$1.4600$ & $0.60595$ & $0.69104$ \\
&CPRRW& $4.3334$ & $1.3238$ & $0.62354$ & $0.71060$ & $4.5520$ &
$1.4475$ & $0.60766$ & $0.69197$ \\
\multicolumn{1}{ c }{{$\Sigma_u^-(2S)$}} &JKM& $4.8243$ & $1.3266$ &
$0.72821$ & $0.96858$ & $5.0405$ & $1.4600$ & $0.70921$ & $0.94390$ \\
&CPRRW& $4.8198$ & $1.3238$ & $0.72845$ & $0.96951$ & $5.0199$ &
$1.4475$ & $0.71085$ & $0.94576$ \\
\multicolumn{1}{ c }{{$\Pi_u^+(1D)$ \& }}
&JKM& $4.1075$ & $1.3266$ & $0.70874$ & $0.80233$ & $4.3472$ &
$1.4600$ & $0.69318$ & $0.78510$ \\
$\Sigma_u^- (1D)$ &CPRRW& $4.1025$ & $1.3238$ & $0.70897$ & $0.80326$
& $4.3246$ & $1.4475$ & $0.69453$ & $0.78650$ \\
\multicolumn{1}{ c }{{ $\Pi_u^+(2D)$ \& }}
&JKM& $4.4991$ & $1.3266$ & $0.78232$ & $1.0231$ \, & $4.7199$ &
$1.4600$ & $0.76341$ & $0.99931$ \\
$\Sigma_u^- (2D)$&CPRRW& $4.4946$ & $1.3238$ & $0.78261$ & $1.0235$ \,
& $4.6989$ & $1.4475$ & $0.76505$ & $1.0012$ \, \\
\hline
\end{tabular}
\end{table*}

The tables are organized by identifying particular exotic states of
known mass and $J^{PC}$ eigenvalues~\cite{Tanabashi:2018oca} as
reference states for particular levels $\Gamma(nL)$ that contain a
state of the given $J^{PC}$.  For charged states, the $C$ value used
is that of its neutral isospin partner.  Our reference states are:
\begin{equation}
\begin{array}{rll}
\hspace{-1em}
 X(3872):    &  M = 3871.69 \pm 0.17 \ {\rm MeV},    & \ J^{PC} = 1^{++}
\, , \\ \hspace{-1em}
Z_c^-(4430): &  M = 4478^{+15}_{-18} \ {\rm MeV}, & \ J^{PC} = 1^{+-}
\, .
\end{array} \hspace{-0.7em}
\end{equation}
In Tables~\ref{table:BestFit}, \ref{table:Fit2}, and
\ref{table:Fit3}, the fits in the left-hand columns correspond to
choosing $X(3872)$ to be the unique $\Sigma^+_g(1S)$ $1^{++}$ state,
one of the two $1^{++}$ states in $\Sigma^+_g(1D)$, and one of the two
$1^{++}$ states in $\Pi^+_u(1P)$-$\Sigma^-_u(1P)$, respectively (these
being the only $n \! = \! 1$, $J^{PC} \! = \! 1^{++}$ states in
Table~\ref{table:States}).  The fits in the right-hand columns
correspond to choosing $Z_c(4430)$ to be one of the two $1^{+-}$
states in $\Sigma^+_g(2S)$, one of the two $1^{+-}$ states in
$\Sigma^+_g(2D)$, and one of the three $1^{+-}$ states in
$\Pi^+_u(2P)$-$\Sigma^-_u(2P)$, respectively (these being the only $n
\! = \! 2$, $J^{PC} \! = \! 1^{+-}$ states in
Table~\ref{table:States}).  The value of $m_\de$ is obtained from
these fits, and is used for all other states in the spectrum of
Tables~\ref{table:BestFit}, \ref{table:Fit2}, and
\ref{table:Fit3}. Also calculated in the tables are values of the
typical length scales $\langle 1/r \rangle^{-1}$ and $\langle r
\rangle$ for the states; as described in Ref.~\cite{Hutson:1994} and
in Appendix~\ref{sec:Comp}, expectation values can be calculated using
the same procedure as one uses to compute eigenvalues, without the
need to generate explicit eigenfunctions.

The choice of $X(3872)$ as an $n \! = \! 1$ state and $Z_c(4430)$ as
an $n \! = \! 2$ state is not logically necessary.  However, if
$X(3872)$ is chosen as the lowest $n \! = \! 2$ state
[$\Sigma^+_g(2S)$], then the mass of the $\Sigma^+_g(1S)$ state is
about $3270$~MeV, which lies squarely in the region of conventional
charmonium between $J/\psi$ and $\chi^{\vphantom\dagger}_{c0}(1P)$.
Such a state with any of the $\Sigma^+_g(1S)$ quantum numbers given in
Table~\ref{table:States} would certainly have been discovered decades
ago, thus rendering the $n \! = \! 2$ assignment untenable.  Similar
comments apply to the $\Sigma^+_g(1S)$ states predicted in
Tables~\ref{table:Fit2} and \ref{table:Fit3}; indeed, the fit of
Table~\ref{table:Fit2} fixing $X(3872)$ to $\Sigma^+_g(1D)$ and the
fit of Table~\ref{table:Fit3} fixing $Z_c(4430)$ to
$\Pi^+_u(2P)$-$\Sigma^-_u(2P)$ produce values $m_\delta \! \simeq \!
1.5 \ {\rm GeV} \! \simeq m_c$, which means that these portions of the
tables effectively reproduce the conventional charmonium spectrum plus
its lowest hybrids, once $\de\bde$ are replaced with $\bar c c$.  The
other fits in Tables~\ref{table:Fit2} and \ref{table:Fit3} predict
exotic $\Sigma^+_g(1S)$ states that lie below the open-charm threshold
at some distance from the conventional charmonium states that are
known to be the only ones populating this region, and hence again
produce conflicts with observation.

As for the $Z_c(4430)$ as an $n \! = \! 1$ state, the assignments
$\Sigma^+_g(1D)$ or $\Pi^+_u(1P)$-$\Sigma^-_u(1P)$ (not tabulated) are
logically possible, but they again lead to a mass for $\Sigma^+_g(1S)$
that is too small (3815 and 3450~MeV, respectively) compared to data,
and also a $\Sigma^+_g(2S)$ mass (4390 and 4045~MeV, respectively)
that, combined with the $\Sigma^+_g(1S)$ states, would generate a
total of at least four $1^{+-}$ states lying below the $Z_c(4430)$.
Experimentally, only two candidates [$Z_c(3900)$ and $Z_c(4200)$] have
confirmed $J^{PC} \! = \! 1^{+-}$, but even the {\em existence\/} of
the $Z_c(4200)$ remains unconfirmed.  In contrast, the existence of
the $Z_c(4020)$ is confirmed, and it is widely expected to be
$1^{+-}$, but only its $C \! = \! -$ quantum number has been
confirmed.  As discussed previously, the assignment of $Z_c(4430)$ to
an $n \!  = \! 1$ state is problematic due to the prediction of
numerous unseen light exotic states, although the choice of
$Z_c(4430)$ as $\Sigma^+_g(1D)$ is not yet definitively excluded,
particularly if a pair of $1^{+-}$ exotics near 4390~MeV [from
$\Sigma^+_g(2S)$] is observed.

One unique level assignment appears to work particularly well with
observation: In Table~\ref{table:BestFit}, $X(3872)$ and $Z_c(4430)$
are identified as states in the multiplets $\Sigma^+_g(1S)$ and
$\Sigma^+_g(2S)$, respectively.  Table~\ref{table:BestFit} exhibits
the prominent feature that its left- and right-hand fits give almost
identical results, supporting the mutual consistency of the chosen
assignments.  The diquark mass obtained is $m_\de \! = \! 1874.7$~MeV
(JKM), 1853.5~MeV (CPRRW), or 1904.2~MeV (BGS), comparing well with
the estimate 1860~MeV used in Ref.~\cite{Brodsky:2014xia}.  The lowest
levels in order of increasing mass are $\Sigma^+_g(1S)$,
$\Sigma^+_g(1P)$, $\Sigma^+_g(2S)$, $\Sigma^+_g(1D)$,
$\Sigma^+_g(2P)$, $\Sigma^+_g(3S)$ (not tabulated,
$\approx$~4890~MeV), $\Pi^+_u(1P)$-$\Sigma^-_u(1P)$, and
$\Sigma^+_g(2D)$.  In particular, only one of the BO potentials beyond
$\Sigma^+_g$ is represented among the lowest states, and even then, it
is in the $6^{\rm th}$ excited level.  What one might call the
``hybrid'' exotic levels begin $\sim$~1~GeV above the $\Sigma^+_g(1S)$
states, just as for hybrid charmonium~\cite{Berwein:2015vca}.  The
lowest observed states in this assignment are therefore
``quark-model'' $\de$-$\bde$ states, in that the gluonic field does
not contribute to the valence spin-parity quantum number.  The states
enumerated in Ref.~\cite{Maiani:2014aja} are included in this list,
although their order and spacing as determined here depends
intrinsically upon the calculated BO potential $\Sigma^+_g(r)$.

The particular mass eigenvalues calculated here apply to all states in
each multiplet $\Gamma(nL)$ listed in Table~\ref{table:States}, which
are degenerate at this level of the calculation: Fine-structure spin-
(and isospin-) dependent corrections are therefore neglected here.
However, one may estimate the magnitude of these mass splittings by
examining those for conventional charmonium.  Note first that
$m_{J/\psi} \! - \!  m_{\eta_c} \! = \!  113$~MeV and
$m_{\chi^{\vphantom\dagger}_{c2}(1P)} \! - \!
m_{\chi^{\vphantom\dagger}_{c0}(1P)} \! = \! 141$~MeV, and the
charmonium fine-structure splittings tend to decrease somewhat with
higher excitation number.  These splittings arise from spin-spin,
spin-orbit, and tensor $c\bar c$ operators all proportional to
$1/m_c^2$.  The corresponding coefficient in the dynamical diquark
model is slightly smaller ($1/m_{\de}^2$), but the diquark spins can
be as large as 1 (compared with $\frac 1 2$ for $c$ or $\bar c$), so
that the spin-operator expectation values in the numerator can be
substantially larger.  Based upon this reasoning, we crudely (but
conservatively) estimate the largest fine-structure splittings in each
exotic multiplets to be $\sim \!  150$~MeV; the true value might turn
out to be even larger, but then one runs the risk of producing
multiple overlapping bands of exotic states, resulting in diminished
model predictivity.  Since the $X(3872)$ appears to be the lowest
exotic candidate, Table~\ref{table:States} then predicts bands of
exotic states in the ranges of approximately 3900--4050~MeV
[$\Sigma^+_g(1S)$], 4220--4370~MeV [$\Sigma^+_g(1P)$], overlapping
bands 4480--4630~MeV [$\Sigma^+_g(2S)$] and 4570--4720~MeV
[$\Sigma^+_g(1D)$], and 4750--4900~MeV [$\Sigma^+_g(2P)$].  Since the
heaviest charmoniumlike state currently observed is $X(4700)$, we do
not analyze the higher levels in any further detail.

In fact, $X(4700)$ and several of the other exotic candidates
[$Y(4140)$, $Y(4274)$, $X(4350)$, $X(4500)$] have only been observed
as resonances decaying to $J/\psi \, \phi$, which makes them good
candidates for $c \bar c s\bar s$ states~\cite{Lebed:2016yvr}, and if
so, then they do not belong to the current analysis; instead, they
would appear as part of an identical analysis using heavier $(cs)$,
rather than $(cu)$ or $(cd)$, diquarks.

Perhaps the most evident pattern among the spectrum of charmoniumlike
exotic bosons~\cite{Lebed:2016hpi}, once the five states listed above
are removed, is the appearance of fairly well-separated clusters: one
between the $X(3872)$ and at least as high as $Z_c(4020)$ [and
possibly as high as the less well-characterized states $Z_c(4050)$ and
$Z_c(4055)$]; another from $Z_c(4200)$ to $Y(4390)$ [and possibly as
low as the less well-characterized $X(4160)$]; the $Z_c(4430)$ by
itself; and the $1^{--}$ states $X(4630)$ and $Y(4660)$.  Even among
the five $c\bar c s\bar s$ candidates, only $Y(4140)$ appears to lie
starkly outside this band structure.  The recently
observed~\cite{Aaij:2018bla} (at $> \!  3\sigma$) $\eta_c \pi^-$
resonance $Z_c(4100)$ also appears to fall into this gap.

Based upon these observations, our central hypothesis is that the
states in the mass range 3872--4055~MeV are $\Sigma^+_g(1S)$ states,
those in 4160--4390~MeV are $\Sigma^+_g(1P)$ states, $Z_c(4430)$ is a
$\Sigma^+_g(2S)$ state, and $X(4630)$ and $Y(4660)$ are
$\Sigma^+_g(2P)$ states.  Note that these bands consist of states with
a single parity: $+$, $-$, $+$, and $-$, respectively.  No known
states need to be assigned to $\Sigma^+_g(1D)$ ($P \! = \! +$).

We now confront this hypothesis with the full data set.  The
charmoniumlike states for which the $J^{PC}$ quantum numbers are
either unambiguously determined or ``favored''
experimentally~\cite{Tanabashi:2018oca} are listed in
Table~\ref{table:JPC}.
\begin{table}
  \caption{Charmoniumlike exotic candidates with experimentally
  determined $J^{PC}$ quantum numbers (both unambiguous and
  ``favored'').}
\label{table:JPC}
\begin{center}
\begin{tabular}{ll}
  \hline\hline
$0^{++^{\vphantom\dagger}}$ & $X(3915)$, $X(4500)$, $X(4700)$ \\
$0^{--}$ & $Z_c^0(4240)$ \\
$1^{--}$ & $Y(4008)$, $Y(4220)$, $Y(4260)$, $Y(4360)$, $Y(4390)$, \\
         & $X(4630)$, $Y(4660)$ \\
$1^{++}$ & $X(3872)$, $Y(4140)$, $Y(4274)$ \\
$1^{+-}$ & $Z_c^0(3900)$, $Z_c^0(4200)$, $Z_c^0(4430)$ \\
$0^+$ \!\!\! or $1^-$ & $Z_c^0(4100)$ \\
${\frac 3 2}^\pm_{\vphantom\dagger}$, ${\frac 5 2}^\mp$ & $P_c(4380)$,
$P_c(4440)$-$P_c(4457)$ \\
  \hline\hline
\end{tabular}
\end{center}
\end{table}
In addition, some dispute remains that the $X(3915)$ could be a
$2^{++}$ state~\cite{Zhou:2015uva} [possibly the same as the
conventional charmonium $\chi^{\vphantom\dagger}_{c2}(2P)$], and the
recently observed
$\chi^{\vphantom\dagger}_{c0}(3860)$~\cite{Chilikin:2017evr} has
properties consistent with being the conventional $0^{++}$
$\chi^{\vphantom\dagger}_{c0}(2P)$, but the existence of this state
has not yet been confirmed.  Charged~\cite{Ablikim:2017oaf} and
neutral~\cite{Ablikim:2017aji} $1^+$ structures from 4032--4038~MeV
are not included because of their uncertain nature.  Lastly, several
states (confirmed and unconfirmed) have unknown $J^P$ but known $C$
eigenvalues~\cite{Tanabashi:2018oca}: $Z_c^0(4020)$ (noted above) and
$Z_c^0(4055)$ are $C \! = \! -$, while $Z_c^0(4050)$, $Z_c^0(4250)$,
$X(4350)$ are $C \! = \! +$.

Under our hypothesis, in $\Sigma^+_g(1S)$ the $X(3872)$ is the sole
$1^{++}$ state, and $X(3915)$ is one of two $0^{++}$ states (although
it could instead be the $c\bar c s\bar s$ ground
state~\cite{Lebed:2016yvr}), with the other $0^{++}$ state possibly
being $\chi^{\vphantom\dagger}_{c0}(3860)$ [instead of the $c\bar c$
$\chi^{\vphantom\dagger}_{c0}(2P)$ assignment], or even possibly (if
fine-structure splittings turn out to be large in this case and $J^P
\! = \! 0^+$ is confirmed) $Z_c^0(4100)$.  The two $1^{+-}$ states are
$Z_c^0(3900)$ and $Z_c^0(4020)$.  As for $2^{++}$, the state
$\chi^{\vphantom\dagger}_{c2}(3930)$ has still not been confirmed as
the $c\bar c$ $\chi^{\vphantom\dagger}_{c2}(2P)$, making it a
potential candidate to complete the $\Sigma^+_g(1S)$ multiplet.  The
states $X(3940)$, $Z_c^0(4050)$, and $Z_c^0(4055)$ are also potential
members (noting the known $C \! = \! +, -$ eigenvalues, respectively,
of the latter two).  The existence of the only other state claimed in
this range, the $1^{--}$ $Y(4008)$~\cite{Yuan:2007sj}, is being
challenged by increasingly adverse evidence, and the state may
disappear completely with newer data and analysis.  In fact,
$Y(4008)$ has $P \! = \! -$ and thus would not fit into
$\Sigma^+_g(1S)$, which represents a success of the model: All
hidden-charm exotics below $4100$~MeV are predicted to have positive
parity.

The $\Sigma^+_g(1P)$ states all have $P \! = \! -$, and indeed,
$Y(4220)$, $Y(4360)$, and (with a small stretch of the band mass
range) $Y(4390)$ fit into this multiplet.  The multiplet actually
contains a fourth $1^{--}$ state, but note that the famous $1^{--}$
$Y(4260)$ may actually be a composite of the other $1^{--}$
states~\cite{BESIII:2016adj}.  The $0^{--}$ $Z_c^0(4240)$ is
especially notable because, if confirmed, it is the lightest state
with exotic $J^{PC}$ ({\it i.e.}, not allowed for conventional $q\bar
q$ mesons).  The $\Sigma^+_g(1P)$ also allows for a pair of $1^{-+}$
($J^{PC}$-exotic) states.  The remaining unassigned states of
$\Sigma^+_g(1P)$ are two each of $0^{-+}$, $2^{-+}$, and $2^{--}$, and
one $3^{--}$.  The unassigned observed states in this mass range are
$X(4160)$, $Y(4274)$ and $X(4350)$ (both identified with $c\bar c
s\bar s$ above), $Z_c^0(4200)$, and $Z_c^0(4250)$.  Of these,
$Z_c^0(4200)$ is problematic because it is a $1^{+-}$ ($P \! = \! +$)
state, but again, its existence remains unconfirmed.

The only clear candidate in the $\Sigma^+_g(2S)$ multiplet is
$Z_c^0(4430)$, although $X(4500)$ [and possibly $X(4700)$, if the
allowed mass range for the band is stretched] are the two potential
$0^{++}$ members; but again, they have been suggested as
$c\bar c s\bar s$ states.  $X(4700)$ can also fit naturally into
$\Sigma^+_g(1D)$.

Finally, the $1^{--}$ $X(4630)$ and $Y(4660)$ states fit into
$\Sigma^+_g(2P)$, assuming the lower bound of the mass range (given
above as 4750--4900~MeV) can be stretched downward slightly.  In fact,
the greatest difficulty of our full level assignment is the tension
between the $Y(4390)$-$X(4630)$ mass difference ($\sim \! 240$~MeV)
and the multiplet $\Sigma^+_g(2P)$-$\Sigma^+_g(1P)$ mass difference
($\sim \!  460$~MeV).  If one supposes that the $Y(4390)$ lies at the
top of the $\Sigma^+_g(1P)$ multiplet and $X(4630)$ lies at the bottom
of the $\Sigma^+_g(2P)$ multiplet, then the assignment remains
sensible.  Clearly a full analysis of fine-structure splittings will
be necessary to assess the fate of this assumption.

We now turn to the hidden-charm pentaquarks, where until recently only
the two states $P_c(4380)$ and $P_c(4450)$ were observed.  Since they
have opposite parities, these states must belong to distinct BO
potential multiplets.  The most recent LHCb
measurements~\cite{Aaij:2019vzc} now resolve the $P_c(4450)$ as two
states, $P_c(4440)$ and $P_c(4457)$, of which presumably at least one
carries opposite parity to the $P_c(4380)$.  In addition, an entirely
new state $P_c(4312)$ has been observed.  While one expects the
diquark mass $m_\de$ to assume the same value as that appearing in
the best fits to the hidden-charm tetraquarks
(Table~\ref{table:BestFit}), the triquark mass $m_\bt$ may be freely
adjusted to fix one of the masses.\footnote{In comparison, the naive
calculation of Ref.~\cite{Lebed:2015tna} assumed $m_\bt \! = \!
m_{\Lambda_c} \! = \! 2.286$~MeV.}  Such a fit, assuming that
$P_c(4450)$ (using the old value~\cite{Tanabashi:2018oca}) is either
the $J^P \! = \!  {\frac{5}{2}}^+$ or ${\frac{3}{2}}^+$ state in
$\Sigma^+(1P)$, is presented in Table~\ref{table:FitPenta}.  However,
then the $P_c(4380)$ must have $J^P \! = \! {\frac{3}{2}}^-$ or
${\frac{5}{2}}^-$.  The latter assignment places it in the higher-mass
$\Sigma^+(1D)$ level (therefore excluded), while the former assignment
places it in the $\Sigma^+(1S)$ level, 370~MeV lower (in contrast with
the observed mass splitting 4450$-$4380$=$70~MeV).  Such a huge mass
difference appears to be impossible to accommodate simply by using
fine-structure effects of a natural size that place $P_c(4380)$ at the
top of its band and $P_c(4450)$ at the bottom of its band.

\begin{table}[ht]
\caption{Mass eigenvalues $M$ (in GeV) for hidden-charm dynamical
diquark-triquark states that are eigenstates of the indicated BO
potentials corresponding to quantum numbers $nL$, for given diquark
$m_{\delta}$ and triquark $m_{\bar\theta}$ masses (in GeV).  The
particular form of the BO potential used is that given by lattice
simulations JKM~\cite{Morningstar:2019} or
CPRRW~\cite{Capitani:2018rox}, or (for the $\Sigma^+_g$ potential) by
the Cornell potential fit to conventional charmonium
BGS~\cite{Barnes:2005pb}.  Also predicted are the corresponding
expectation values for the length scales ${\langle 1/r \rangle}^{-1}$
and $\langle r \rangle$ (in fm).  Fixing $m_{\delta}$  from the
corresponding simulation in Table~\ref{table:BestFit} and fixing to
the experimental mass of $P_c(4450)$ (as indicated by boldface)
predicts $m_{\bar\theta}$ and the whole spectrum.}
\label{table:FitPenta}
\setlength{\extrarowheight}{0.8ex}
\begin{center}
\begin{tabular}{ c c c c c c c }
\hline\hline
\multicolumn{2}{c}{\multirow{2}{*}{}}&\multicolumn{5}{c}{ }\\
\cline{3-4}\hline
BO states & Potential & $M$ & $m_{\delta}$&$m_{\bar{\theta}}$&
${\langle}{1/r}{\rangle}^{-1}$&${\langle}{r}{\rangle}$ \\
\hline
\multicolumn{1}{ c }{{$\Sigma_g^+(1S)$}}
&JKM& $4.0788$ & $1.8747$ & $2.0987$ & $0.26545$ & $0.35646$   \\
&CPRRW& $4.0821$ & $1.8535$ & $2.0800$ & $0.26847$ & $0.36019$   \\
&BGS& $3.9718$ & $1.9402$ & $2.0527$ & $0.20939$ & $0.29732$ \\
\multicolumn{1}{ c }{{$\Sigma_g^+(2S)$}}
&JKM& $4.6430$ & $1.8747$ & $2.0987$ & $0.41816$ & $0.67707$   \\
&CPRRW& $4.6431$ & $1.8535$ & $2.0800$ & $0.42161$ & $0.68219$   \\
&BGS& $4.5682$ & $1.9402$ & $2.0527$ & $0.42091$ & $0.68871$ \\
\multicolumn{1}{ c }{{$\Sigma_g^+(1P)$}}
&JKM& $\textbf{4.4498}$ & $1.8747$ & $2.0987$ & $0.47967$ & $0.55484$
\\
&CPRRW& $\textbf{4.4498}$ & $1.8535$ & $2.0800$ & $0.48356$ &
$0.55903$  \\
&BGS& $\textbf{4.4498}$ & $1.9402$ & $2.0527$ & $0.47727$ & $0.56066$
\\
\multicolumn{1}{ c }{{$\Sigma_g^+(2P)$}}
&JKM& $4.9094$ & $1.8747$ & $2.0987$ & $0.61225$ & $0.82655$ \\
&CPRRW& $4.9078$ & $1.8535$ & $2.0800$ & $0.61665$ & $0.83237$   \\
&BGS& $4.8421$ & $1.9402$ & $2.0527$ & $0.64586$ & $0.88662$ \\
\multicolumn{1}{ c }{{$\Sigma_g^+(1D)$}}
&JKM& $4.7317$ & $1.8747$ & $2.0987$ & $0.65125$ & $0.71735$ \\
&CPRRW& $4.7303$ & $1.8535$ & $2.0800$ & $0.65592$ & $0.72224$   \\
&BGS& $4.7170$ & $1.9402$ & $2.0527$ & $0.68984$ & $0.76462$ \\
\multicolumn{1}{ c }{{$\Sigma_g^+(2D)$}}
&JKM& $5.1405$ & $1.8747$ & $2.0987$ & $0.77137$ & $0.96182$ \\
&CPRRW& $5.1379$ & $1.8535$ & $2.0800$ & $0.77667$ & $0.96811$   \\
&BGS& $5.0481$ & $1.9402$ & $2.0527$ & $0.83674$ & $1.0533\;$ \\
\multicolumn{1}{ c }{{$\Pi_u^+(1P)$ \& }}
&JKM& $5.1320$ & $1.8747$ & $2.0987$ & $0.44420$ & $0.56275$ \\
$\Sigma_u^- (1P)$&CPRRW& $5.0971$ & $1.8535$ & $2.0800$ & $0.44110$ &
$0.55763$   \\
\multicolumn{1}{ c }{{$\Pi_u^-(1P)$}}
&JKM& $5.2092$ & $1.8747$ & $2.0987$ & $0.64219$ & $0.72457$ \\
&CPRRW& $5.2417$ & $1.8535$ & $2.0800$ & $0.65372$ & $0.73737$   \\
\multicolumn{1}{ c }{{$\Sigma_u^-(1S)$}}
&JKM& $5.5224$ & $1.8747$ & $2.0987$ & $0.58197$ & $0.62794$ \\
&CPRRW& $5.5005$ & $1.8535$ & $2.0800$ & $0.55634$ & $0.63400$   \\
\multicolumn{1}{ c }{{$\Pi_u^+(1D)$ \& }}
&JKM& $5.3139$ & $1.8747$ & $2.0987$ & $0.65592$ & $0.74366$ \\
$\Sigma_u^- (1D)$&CPRRW& $5.2847$ & $1.8535$ & $2.0800$ & $0.64810$ &
$0.73365$   \\
\hline
\end{tabular}
\end{center}
\end{table}

Much more natural for matching with the known spectroscopy, since the
$\Sigma^+(2S)$-$\Sigma^+(1P)$ multiplet-average mass difference is
calculated to be only $\simeq \! 200$~MeV, is to identify $P_c(4380)$
as the ${\frac{5}{2}}^+$ state in $\Sigma^+(1P)$, fix to the mass of
the $P_c(4312)$ as the bottom of the $\Sigma^+(1P)$, and identify
$P_c(4440)$, $P_c(4457)$ as belonging to $\Sigma^+(2S)$, one of them
being its $J^P \! = \!  {\frac{3}{2}}^-$ state.
Table~\ref{table:FitPenta2} presents this rather satisfactory fit.
Clearly, measuring $P$ for any of the observed states would
distinguish these scenarios.  In fits with this level assignment, we
notably find $m_\bt \! \approx \! 1.93$~GeV, which is only slightly
larger than $m_\de$.  We also predict $\Sigma^+(1S)$ hidden-charm
pentaquark ground states in this fit to lie near 3940~MeV; such
states would be stable against decay to $J/\psi \, N$ and possibly
even to $\eta_c N$.  They would decay through annihilation of the
$c\bar c$ pair to light hadrons plus a nucleon, and would have narrow
widths, comparable to that of $\eta_c(1S)$ [$O(10)$ rather than
$O(100)$~MeV].

\begin{table}[ht]
\caption{As in Table~\ref{table:FitPenta}, except now assuming that
$P_c(4312)$ is a $\Sigma_g^+(1P)$ state (as indicated by boldface) in
order to fix the full spectrum.}
\label{table:FitPenta2}
\setlength{\extrarowheight}{0.8ex}
\begin{center}
\begin{tabular}{c c c c c c c } 
\hline\hline
\multicolumn{2}{c}{\multirow{2}{*}{}}&\multicolumn{5}{c}{ }\\
\cline{3-4}
\hline
BO states & Potential & $M$ & $m_{\delta}$&$m_{\bar{\theta}}$&
${\langle}{1/r}{\rangle}^{-1}$&${\langle}{r}{\rangle}$ \\
\hline
\multicolumn{1}{ c }{{$\Sigma_g^+(1S)$}}
&JKM& $3.9385$ & $1.8747$ & $1.9478$ & $0.26975$ & $0.36205$   \\
&CPRRW& $3.9419$ & $1.8535$ & $1.9291$ & $0.27282$ & $0.36578$   \\
&BGS& $3.8375$ & $1.9402$ & $1.9014$ & $0.21497$ & $0.30478$ \\
\multicolumn{1}{ c }{{$\Sigma_g^+(2S)$}}
&JKM& $4.5078$ & $1.8747$ & $1.9478$ & $0.42390$ & $0.68615$   \\
&CPRRW& $4.5080$ & $1.8535$ & $1.9291$ & $0.42740$ & $0.69151$   \\
&BGS& $4.4329$ & $1.9402$ & $1.9014$ & $0.42813$ & $0.70082$ \\
\multicolumn{1}{ c }{{$\Sigma_g^+(1P)$}}
&JKM& $\textbf{4.3119}$ & $1.8747$ & $1.9478$ & $0.48619$ & $0.56229$
\\
&CPRRW& $\textbf{4.3119}$ & $1.8535$ & $1.9291$ & $0.49013$ &
$0.56671$   \\
&BGS& $\textbf{4.3119}$ & $1.9402$ & $1.9014$ & $0.48591$ & $0.57067$
\\
\multicolumn{1}{ c }{{$\Sigma_g^+(2P)$}}
&JKM& $4.7764$ & $1.8747$ & $1.9478$ & $0.62012$ & $0.83726$   \\
&CPRRW& $4.7749$ & $1.8535$ & $1.9291$ & $0.62463$ & $0.84331$   \\
&BGS& $4.7073$ & $1.9402$ & $1.9014$ & $0.65602$ & $0.90036$ \\
\multicolumn{1}{ c }{{$\Sigma_g^+(1D)$}}
&JKM& $4.5965$ & $1.8747$ & $1.9478$ & $0.65975$ & $0.72643$   \\
&CPRRW& $4.5952$ & $1.8535$ & $1.9291$ & $0.66444$ & $0.73155$   \\
&BGS& $4.5802$ & $1.9402$ & $1.9014$ & $0.70076$ & $0.77649$ \\
\multicolumn{1}{ c }{{$\Sigma_g^+(2D)$}}
&JKM& $5.0099$ & $1.8747$ & $1.9478$ & $0.78105$ & $0.97393$   \\
&CPRRW& $5.0074$ & $1.8535$ & $1.9291$ & $0.78648$ & $0.98045$   \\
&BGS& $4.9146$ & $1.9402$ & $1.9014$ & $0.84914$ & $1.0687$ \\
\multicolumn{1}{ c }{{$\Pi_u^+(1P)$ \& }}
&JKM& $4.9860$ & $1.8747$ & $1.9478$ & $0.44758$ & $0.56718$   \\
$\Sigma_u^- (1P)$&CPRRW& $4.9514$ & $1.8535$ & $1.9291$ & $0.44434$ &
$0.56205$   \\
\multicolumn{1}{ c }{{$\Pi_u^-(1P)$}}
&JKM& $5.0982$ & $1.8747$ & $1.9478$ & $0.65924$ & $0.74413$   \\
&CPRRW& $5.0661$ & $1.8535$ & $1.9291$ & $0.64771$ & $0.73109$   \\
\multicolumn{1}{ c }{{$\Sigma_u^-(1S)$}}
&JKM& $5.3816$ & $1.8747$ & $1.9478$ & $0.58690$ & $0.63446$   \\
&CPRRW& $5.3597$ & $1.8535$ & $1.9291$ & $0.56217$ & $0.64075$   \\
\multicolumn{1}{ c }{{$\Pi_u^+(1D)$ \& }}
&JKM& $5.1714$ & $1.8747$ & $1.9478$ & $0.66148$ & $0.74995$   \\
$\Sigma_u^- (1D)$&CPRRW& $5.1426$ & $1.8535$ & $1.9291$ & $0.65333$ &
$0.73970$   \\
\hline
\end{tabular}
\end{center}
\end{table}

It should be noted that the fits in
Tables~\ref{table:FitPenta},\ref{table:FitPenta2} discussed in the
previous paragraph use the full ``homonuclear'' BO potentials
determined on the lattice, including the reflection quantum number
$\eta$.  However, this reflection symmetry disappears for the
``heteronuclear'' system.  Nevertheless, the ground-state potential
$\Sigma^+_g$ in the ``homonuclear'' case is well separated from any
other BO potential (in particular, from $\Sigma^+_u$), so that
nothing is lost by using it as the ``heteronuclear'' ground-state
$\Sigma^+$ potential.  For the higher potentials (which we did not use
in the phenomenological analysis), the ``homonuclear'' BO potentials
represent the interactions of a system with two equal masses $2\mu$
but with the same separation parameter $r$.

The final results involve the BO decay selection rules first discussed
for exotics in Refs.~\cite{Braaten:2014ita,Braaten:2014qka} and
obtained for this model in Ref.~\cite{Lebed:2017min}.  These rules
assume not only that the light d.o.f.\ decouple from the heavy $\QQ$
pair, and that they adjust more rapidly in a physical process than the
$\QQ$, but also that the quantum numbers of the $\QQ$ and the light
d.o.f.\ are separately conserved in the decay.  As noted in
Ref.~\cite{Lebed:2017min}, the strictest tests of the BO decay
selection rules occur for single light-hadron decays ($\pi$, $\rho$,
$\omega$, $\phi$) of exotics to conventional charmonium ($\Sigma^+_g$)
states.  Since $Z_c(3900)$ has $J^P \! = \! 1^+$, the observed decay
$Z_c^+(3900) \! \to \! J/\psi \, \pi^+$ requires $\pi^+$ to be in an
even partial wave, and furthermore requiring conservation of
heavy-quark spin [$s_\QQ \!  = \! 1$ for $J/\psi$, and hence also for
$Z_c^+(3900)$], Ref.~\cite{Lebed:2017min} found
$\Pi^+_u(1P)$-$\Sigma^-_u(1P)$ to be the most likely home for
$Z_c(3900)$.  However, as we have seen, the
$\Pi^+_u(1P)$-$\Sigma^-_u(1P)$ level lies $\sim \! 1$~GeV above the
ground-state level $\Sigma^+_g(1S)$, which is untenable for
phenomenology.

The situation with the vector-meson decays is even worse, as the decay
selection rules, when strictly applied as above, require the
introduction of BO potentials beyond those listed in
Table~\ref{table:States}.  Lattice simulations predict these levels to
lie even higher in mass ($> \! 1$~GeV) above $\Sigma^+_g(1S)$.  And in
the pentaquark decays, either $\Pi^+(1D)$ or $\Sigma^-(1S)$, which are
highly excited levels, is given as the favored home for the $P \! = \!
+$ candidate.  The strictest application of the BO decay selection
rules appears to conflict with the known spectroscopy.

The simplest way to resolve such issues is to note that the evidence
for the conservation of $\QQ$ spin in exotics (as discussed in
Ref.~\cite{Lebed:2017min}) is imperfect, meaning that the requirement
of separate conservation of $\QQ$ spin and light d.o.f.\ quantum
numbers, which forced unacceptably high BO potentials to appear, may
also be called into question.  More precisely, in contrast to
conventional quarkonium states, the exotics do not obviously occur in
eigenstates of heavy-quark spin-symmetry.  A better approach, as
suggested by this work, appears to be that of obtaining the
spectroscopy in a robust calculation, and then from the observed
decays identifying the behavior of the states' internal quantum
numbers---as is done for conventional quarkonium transitions.

\section{Approximations of The Model}
\label{sec:Approx}
We have modeled mass eigenstates formed from a $\de$-$\bde$ (or
$\bt$-$\de$) pair of sufficient relative momentum to create between
them a color flux tube of substantial spatial extent, but not so large
as to induce immediate fragmentation of the flux tube.  In this
section, almost all the comments applied to $\de$-$\bde$ systems also
apply to $\bt$-$\de$ systems.

The first and most obvious question is whether the quantized states
of such a configuration of {\em dynamical\/} origin are best
described in terms of the {\em static\/} configuration provided by
lattice simulations.  We have argued that the transition from the
former to the latter paradigm is facilitated by the WKB
approximation, specifically by the enhancement of the amplitude when
the $\de$-$\bde$ system approaches its classical turning point.

The original result~\cite{Brodsky:2014xia} $r \! = \! 1.16$~fm for the
$Z_c^-(4430)$ spatial extent, at which point the $\de$-$\bde$ pair in
the process $B^0 \! \to \!  (\psi(2S) \pi^- ) K^+$ comes completely to
rest, is only slightly smaller than the 1.224(15)~fm lattice
calculation of the string-breaking distance very recently presented in
Ref.~\cite{Bulava:2019iut}.  However, the result of
Ref.~\cite{Brodsky:2014xia} explicitly depends upon available phase
space for the $\de$-$\bde$ pair, and hence upon $m_{B^0}$ and
$m_{K^+}$ (in addition to $m_{\de}$).  Of course, the mass eigenvalue
of a $c\bar c d\bar u$ bound state should depend only upon its
internal dynamics, and not upon the details of the process through
which it is produced; therefore, 1.16~fm should be viewed as a
theoretical maximum for the possible size of the exotic state
$Z_c^-(4430)$, in contrast to the smaller values of $\langle r
\rangle$ computed above.  In particular, the constituents of the
actual bound state should carry nonzero internal kinetic energy,
which the naive calculation ignores.  Since lattice-calculated static
potentials provide the best available {\it ab initio\/} information
on the nature of gluonic fields of finite spatial extent, they
provide the most natural framework for modeling $\de$-$\bde$ bound
states, even ones of dynamical origin.

The next obvious approximation is that this model assumes
(effectively) structureless, pointlike $\de$ quasiparticles.  A true
diquark quasiparticle---setting aside the fact that (like a quark) it
carries nonzero color charge and therefore is a gauge-dependent
object---should have a finite size, comparable to that of a heavy
meson (a few tenths of a fm).  But then, the states obtained above
have a natural size only 2--4 times larger, meaning that the notion
of a tetraquark state with well-separated components comes into
question.  Corrections that probe the robustness of the present
results by including the finite size of diquarks as a perturbation
are planned for subsequent work~\cite{Giron:2019}.

Another consequence of the assumption of structureless diquarks is the
absence of both spin- and isospin-dependent effects.  Each row of
Tables~\ref{table:States} and \ref{table:States2} lists all the
eigenstates of specific quantum numbers $n$ and $L$ for a particular
BO potential $\Gamma$, which are degenerate at this stage of the
calculation.  Inclusion of the requisite {\em fine-structure\/}
corrections is necessary to lift the degeneracies and to produce a
full spectrum of states.  At this juncture, the present numerical
results are identical to those one would obtain by using the methods
of Ref.~\cite{Berwein:2015vca} for hybrid mesons, except with the
heavy-quark mass replaced by the somewhat heavier diquark/triquark
mass, and with the full quantum numbers for the states obtained only
after including the light-quark spins and isospins.

The most important fine-structure corrections identified here fall
into two categories: First are the spin-spin interactions within each
of $\de$ and $\bde$ (or $\bt$); their importance in understanding the
fine structure of the exotics spectrum was first emphasized in
Ref.~\cite{Maiani:2014aja}.  Second, since each of $\de$ and $\bde$
contains a light quark (or two for $\bt$), one expects in general
long-distance spin- {\em and\/} isospin-dependent corrections to
modify the spectrum; without the latter, the $\de$-$\bde$ states
would fall into degenerate $u\bar u, u\bar d, d\bar u, d\bar d$
quartets rather than into the experimentally observed $I \! = \! 0$
singlets and $I \! = \! 1$ triplets.  Were the exotic states instead
composed of molecules of two isospin-carrying hadrons, a natural
differentiation in the spectrum based upon isospin would
arise~\cite{Cleven:2015era}, {\it e.g.}, through distinct couplings
of the hadrons via (long-distance) $\pi$ exchange {\it vs}.\ $\eta$
exchange.

In contrast, the dominant interaction in the dynamical diquark model
between $\de$ and $\bde$ (or $\bt$) occurs through the color-adjoint
flux tube.  However, even in this case one can identify
isospin-dependent interactions through the exchange of colored
$\pi$-like quasiparticles, owing to a variant of the Nambu-Goldstone
theorem of chiral-symmetry breaking originally discussed in the
context of {\it color-flavor locking\/}~\cite{Alford:1998mk}.  Thus,
one expects modifications to the spectrum arising from long-distance
spin- and isospin-dependent interactions.  As suggested in
Ref.~\cite{Lebed:2017min}, lattice simulations in which the heavy,
static sources are assigned quantum numbers of not only color and
spin but also isospin would have excellent investigative value for
this scenario.  Incorporation of both the diquark-internal and
long-distance $\de$-$\bde$ fine-structure corrections is planned for
the next refinement of model calculations~\cite{Giron:2019}.

Absent in the discussion up to this point is perhaps the most
consequential of all corrections for the meson sector: For most
$J^{PC}$ quantum numbers, the physical heavy hidden-flavor meson
spectrum likely contains not only possible $\de$-$\bde$ states, but
also ordinary $\QQ$ quarkonium, as well as $\bar Q Q g$ hybrids, in
addition to molecules of heavy-meson pairs. {\it Coupled-channel\/}
mixing effects to include all of these states can have a profound
effect on the observed spectrum.  For example, a commonly held view
in the field~\cite{Lebed:2016hpi,Chen:2016qju,Hosaka:2016pey,
Esposito:2016noz,Guo:2017jvc,Ali:2017jda,Olsen:2017bmm,
Karliner:2017qhf,Yuan:2018inv} is that the peculiar properties of the
$X(3872)$---particularly, its extreme closeness to the
$D^0$-$\bar D^{*0}$ threshold, its small width, and its substantial
collider prompt-production rate---can be explained by $X(3872)$ being
an admixture of a $D^0$-$\bar D^{*0}$ molecule and the yet-unseen
$\chi_{c1}(2P)$ charmonium state.  The addition of $\de$-$\bde$
states clearly makes the complete spectrum all the more rich and
complex.  At this stage of the dynamical diquark study, we do not
attempt to address this intricate and deeply interesting problem.

\section{Discussion and Conclusions}
\label{sec:Concl}

We have produced the first numerical predictions of the dynamical
diquark model, which is the application of the Born-Oppenheimer (BO)
approximation to the dynamical diquark {\em picture}.  In turn, this
picture describes a multiquark exotic state as a system of a compact
diquark $\de$ and antidiquark $\bde$ for a tetraquark (or triquark
$\bt$ for a pentaquark) interacting through a gluonic field of finite
extent.  Using the results of lattice simulations for the lowest BO
potentials, we have obtained the mass eigenvalues of the
corresponding Schr\"{o}dinger equations (both uncoupled and coupled),
and found that all known hidden-charm multiquark exotic states can be
accommodated by the lowest ($\Sigma^+_g$) potential, for which the
gluonic quantum numbers are $J^P \! = \!  0^+$.  In this sense, our
explicit calculations support a type of ``quark-model'' classification
of the lowest multiquark exotics, in which one obtains the tetraquark
or pentaquark quantum numbers by combining $\de$ and $\bde$ (or $\bt$)
quantum numbers and their relative orbital angular momentum, exactly
as one does for $\qq$ mesons.

Each level $\Gamma(nL)$ for each BO potential $\Gamma$ produces a
distinct mass eigenvalue, but differences due to the spin (and
isospin) quantum numbers of the $\de$, $\bde$($\bt$) are ignored in
this calculation, meaning that each mass eigenvalue corresponds to a
degenerate multiplet of states with various $I^G$, $J^{PC}$ quantum
numbers.  We estimated the maximum size of the neglected
fine-structure splittings ($\sim \! 150$~MeV), and found that the
spectrum of tetraquarks should consist of a lowest [$\Sigma^+_g(1S)$]
band, all members of which have $P \! = \! +$, followed by a gap of
about 100~MeV, and then a $\Sigma^+_g(1P)$ band of $P \! = \! -$
states, then another gap and overlapping $\Sigma^+_g(2S)$ and
$\Sigma^+_g(1D)$ bands of $P \! = \! +$ states, and finally a band of
$\Sigma^+_g(2P)$, $P \! = \! -$ states.  The order for pentaquark
states is the same, except that the reflection quantum number ``$g$''
is no longer present, and the $P$ eigenvalues are opposite those for
tetraquarks.  Many higher levels are predicted, but are not yet needed
to accommodate currently observed exotics.

As of the present, the computed band structure with alternating $P$
is supported by the known states, such that $X(3872)$ [$Z_c^-(4430)$]
is a member of the $\Sigma^+_g(1S)$ [$\Sigma^+_g(2S)$] multiplet.
The exceptions are the $1^{++}$ $Y(4140)$, which lies in the first
band gap, but may be a $c\bar c s\bar s$ state and thus fall outside
the current analysis; and the $1^{+-}$ $Z_c^0(4200)$ (if its
existence is confirmed), since it lies in the region of the
$\Sigma^+_g(1P)$, $P \! = \!  -$ band.  For the pentaquarks, the
small $P_c(4457)$-$P_c(4380)$ mass difference is most easily
accommodated by assigning the heavier state to the $2S$ band and the
lighter one to the $1P$ band, leading to the prediction of $1S$-band
hidden-charm pentaquarks that may be stable against decay into
charmed particles.

However, the BO decay selection rules, based upon separate
conservation of heavy quark-antiquark and light degree-of-freedom
quantum numbers in observed decay processes, mandate that known,
low-lying exotics must appear in highly excited BO potentials, and
thus contradict observation.  We propose that the selection rules fail
badly because they are based in part upon assigning the exotics to
heavy-quark spin eigenstates, for which the experimental evidence
appears to be quite mixed.

To develop the model further, one must perform a detailed analysis of
the fine-structure corrections (both spin and isospin dependence) to
determine whether the specific level structure suggested by data is
supported by experiment.  One must also include effects arising from
finite diquark (triquark) sizes.  These refinements will be
implemented in subsequent work to be carried out by this
collaboration.

Additional improvements rely upon, first of all, a reassessment of
lattice simulations: How much does the spin of the heavy sources
($\frac 1 2$ for $c$ and $\bar c$, 0 or 1 for $\de$, $\bde$), or the
light-flavor content in the diquark/triquark case, modify the BO
potentials?  Finally, this work assumes that {\em every\/} exotic
state in the charmoniumlike system is a $\de$-$\bde$ or $\de$-$\bt$
state.  Ignored completely in this analysis are the possibilities
that some of these states are high conventional $c\bar c$, or that
some are genuinely hadronic molecules, or threshold effects, or even
mixtures of these types.  Only a global analysis including
observables such as detailed branching fractions and production
lineshapes can truly disentangle the full spectrum.

\begin{acknowledgments}
  R.F.L.\ was supported by the National Science Foundation (NSF) under
  Grant No.\ PHY-1803912; J.F.G.\ through the Western Alliance to
  Expand Student Opportunities (WAESO) Louis Stokes Alliance for
  Minority Participation Bridge to the Doctorate (LSAMPBD) NSF
  Cooperative Agreement HRD-1702083; and C.T.P.\ through a NASA
  traineeship grant awarded to the Arizona/NASA Space Grant
  Consortium.  The authors also gratefully acknowledge discussions
  with J.M.~Hutson, C.R.~Le~Sueur, M.~Berwein, H.~Martinez, and
  A.~Saurabh.
\end{acknowledgments}
 
\appendix

\section{Born-Oppenheimer Potential Quantum Numbers for a
``Diatomic'' $\QQ$ System}
\label{sec:BOapp}

\begin{figure}[ht!]
\begin{center} \includegraphics[width = \linewidth]{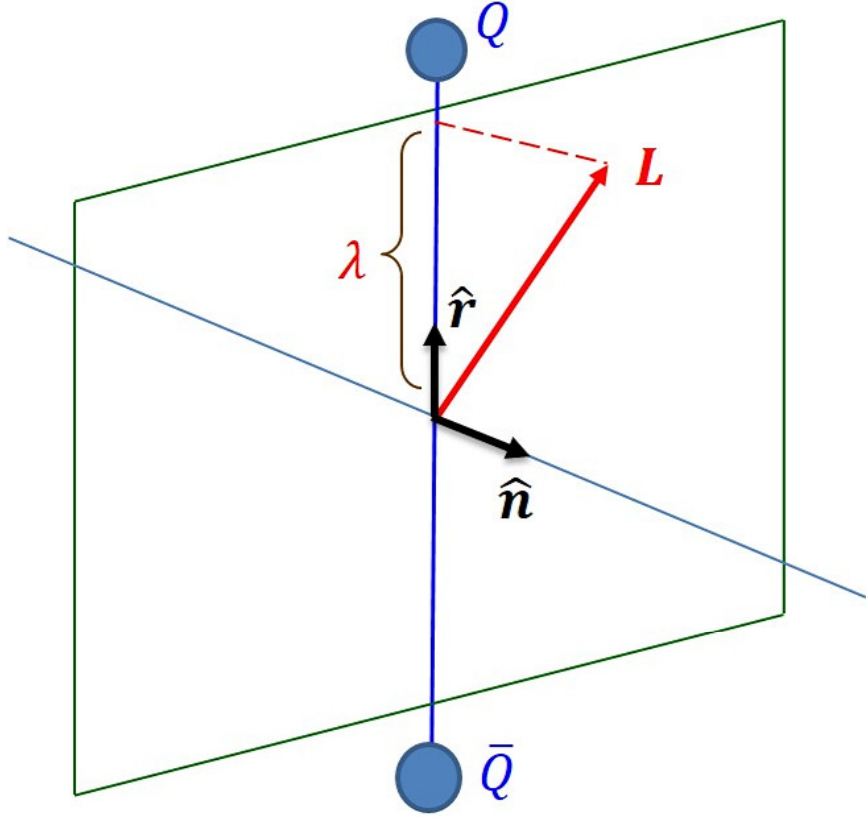}
\caption{Symmetry directions and quantum numbers relevant to a
``diatomic'' $\QQ$ system.  The system is ``homonuclear'' when $Q$
and $\bar Q$, or $\delta$ and $\bar \delta$, are antiparticles of
each other, and ``heteronuclear'' otherwise.
\label{fig:BO_Quantum}}
\end{center}
\end{figure}

We define here the conventional notation for BO potentials used in the
classic Ref.~\cite{Landau:1977}.  A system with two heavy sources has
a relative separation $r$ and a characteristic unit vector $\hat {\bf
r}$ connecting them, as depicted in Fig.~\ref{fig:BO_Quantum}.  In
the $\QQ$ ($\de \bde$, $\de \bt$) case, $\hat {\bf r}$ points from
$\bar Q$ to $Q$ ($\bde$ to $\de$ or $\bt$ to $\de$).  The system may
be ``homonuclear'' (when $Q$ and $\bar Q$, or $\de$ and $\bde$, are
antiparticles of each other), or ``heteronuclear'' otherwise, such as
for $B_c$ multiquark exotics or the {\it triquark}-diquark pentaquark
configuration described in Refs.~\cite{Lebed:2015tna,Lebed:2017min}.
The ``homonuclear'' (``heteronuclear'') system possesses the same
symmetry group $D_{\infty h}$ ($C_{\infty v}$) as a cylinder (cone)
with axis $\hat {\bf r}$.

Since the BO potentials in the two-heavy-source case depend only upon
$r$, the potentials connecting the $\de$-$\bde$ pair can be labeled
by the irreducible representations of $D_{\infty h}$.  The
conventional quantum numbers used~\cite{Landau:1977} are $\Gamma \!
\equiv \! \Lambda^\epsilon_\eta$. $\Lambda$ is an angular momentum
projection, and $\epsilon$ and $\eta$ are inversion parities.  With
reference to Fig.~\ref{fig:BO_Quantum}, one defines a plane
containing the axis $\hat {\bf r}$ and a unit normal $\hat {\bf n}$
to the plane.  Denoting the total angular momentum of the light
d.o.f.---a conserved quantity, thanks to the decoupling in the BO
approximation---as $\bm{J}_{\rm light}$, and the orbital angular
momentum of the heavy d.o.f.\ as $\bm{L}_\QQ$, one obtains the total
orbital angular momentum of the system,
\begin{equation} \label{eq:Ldef}
\bm{L} \equiv \bm{L}_\QQ + \bm{J}_{\rm light} \, .
\end{equation}
Since $\hat {\bm{r}} \! \cdot \! \bm{L}_\QQ = 0$, the axial angular
momentum $\hat {\bm{r}} \! \cdot \! \bm{J}_{\rm light} \! = \hat
{\bm{r}} \! \cdot \! \bm{L}$ of the light d.o.f.\ is a good quantum
number for the whole system, and its eigenvalues are denoted by
$\lambda = 0, \pm 1, \pm 2, \ldots$.  One further notes that
\begin{equation} \label{eq:LambdaMax}
L \ge |\hat {\bm{r}} \cdot \bm{L}| = |\hat {\bm{r}} \cdot
\bm{J}_{\rm light}| = |\lambda| \equiv \Lambda \, .
\end{equation}
Analogous to the use of labels $S,P,D,\ldots$ for the quantum numbers
$L=0,1,2,\ldots$, one denotes potentials with the eigenvalues
$\Lambda=0,1,2,\ldots$ by $\Sigma,\Pi,\Delta,\ldots$.

Reflection $R_{\rm light}$ of the light d.o.f.\ through the plane with
unit normal $\hat {\bf n}$ (which is spatial inversion $P_{\rm light}$
of the light d.o.f., combined with a rotation by $\pi$ radians about
$\hat {\bf n}$ originating from the $\de$-$\bde$ midpoint) transforms
$\lambda \! \to \! -\lambda$.  Since a glance at
Fig.~\ref{fig:BO_Quantum} reveals that the physical system (and hence
its energy) must be invariant under $R_{\rm light}$, one finds that
the energy eigenvalues must be a function only of $\Lambda \!
\equiv \! |\lambda|$.

The eigenvalues of $R_{\rm light}$ itself are denoted $\epsilon \! =
\! \pm 1$.  Strictly speaking, specifying $\epsilon$ is essential
only for $\Sigma$ potentials, for which the $\epsilon \! = \! \pm 1$
states may have distinct energies; however, one may also form
$\epsilon$ eigenstates for (degenerate) $\Lambda \! > \! 0$ BO
potentials in the same way as one uses linear combinations of $f(+x)$
and $f(-x)$ to form  both even and odd functions from an arbitrary
function $f$.  Following this procedure~\cite{Lebed:2017min}, one
obtains BO potentials for all values of $\Lambda^\epsilon$, in which
case one also finds $P_{\rm light} \!  = \! \epsilon (-1)^\Lambda$ in
the ``homonuclear'' case.

Even for the ``homonuclear'' system, a complete inversion of
coordinates of the light d.o.f.\ through the midpoint of the $\QQ$
pair does not by itself produce an equivalent state to the original;
rather, one must also exchange the $\QQ$ pair, or instead, perform
charge conjugation $C_{\rm light}$ upon the light d.o.f.\  Thus, the
full system in the ``homonuclear'' case is physically invariant under
the combination $(CP)_{\rm light}$, and its eigenvalues $\eta \! = \!
+1,-1$ are labeled as $g$, $u$, respectively.

In total, the three eigenvalues $\Lambda^\epsilon_\eta$ completely
specify the irreducible representation of $D_{\infty h}$ for any
``homonuclear diatomic'' system.\footnote{In a true atomic system, the
total electron spin $s$ in the light d.o.f.\ is appended as a
superscript, as in ${}^{2s+1}
\Lambda^\epsilon_\eta$~\cite{Landau:1977}.}  In the ``heteronuclear
diatomic'' case (such as $B_c$ tetraquarks or the triquark-diquark
pentaquark), the $(CP)_{\rm light}$ symmetry is lost (leaving the
symmetry group $C_{\infty v}$), but the good quantum numbers $\Gamma
\! \equiv \! \Lambda^\epsilon$ remain.

\section{Computational Methods}
\label{sec:Comp}

For both $n$-fold coupled and uncoupled reduced radial Schr\"odinger
equations, one must solve a Sturm-Liouville problem of the form
\begin{equation}\label{eqn:sturm_louisville_eqn}
    \left[\mathbb{I}\frac{\diff^2 }{\diff r^2}+\mathbf{Q}(r)\right]
    \! u(r)=0 \, ,
\end{equation}
such that
\begin{equation}\label{eqn:sturm_louisville}
    \mathbf{Q}(r)\equiv\frac{2\mu}{\hbar^2}\left(E \, \mathbb{I}-
    \frac{\hat{\mathbf{L}}^2}{2\mu r^2}-\mathbf{V}(r)\right) \ \
    \textrm{and} \ u(0)=0 \, ,
\end{equation}
where $u$ is an $n$-dimensional column vector, $\mathbb{I}$ is an
$n\times n$ identity matrix, $\hat{\mathbf{L}}^2$ is a matrix
representing the angular momentum operator of the coupled system,
$\mathbf{V}(r)$ is the interaction potential matrix, and $E$ is the
energy eigenvalue of the Hamiltonian.  One now takes the $n$ linearly
independent column-vector solutions to
Eq.~(\ref{eqn:sturm_louisville_eqn}) and concatenates them into an
$n\times n$ matrix $\Umat(r)$. One can then numerically integrate each
$u_{m}(r)$ for $m\in\{1,2,...,n\}$ simultaneously, via the equivalent
problem:
\begin{equation}\label{eqn:johnson_schrod_U}
    \left[\mathbb{I}\frac{\diff^2 }{\diff r^2}+\mathbf{Q}(r)\right]
    \! \Umat(r)=\textbf{0} \, .
\end{equation}

The Sturm-Liouville problem of
Eqs.~(\ref{eqn:sturm_louisville_eqn})--(\ref{eqn:sturm_louisville})
has a solution for any $E\in\mathbb{R}$. Since the probabilistic
interpretation of quantum mechanics requires that solutions to
Eq.~(\ref{eqn:johnson_schrod_U}) must be well-behaved as well as {\em
normalizable}, one must have that the solution asymptotically
approaches zero in the $r\rightarrow \infty$ limit:
\begin{equation}\label{eqn:right_boundary_condition}
    \lim_{r\rightarrow\infty}\Umat(r)\rightarrow \textbf{0} \, .
\end{equation}
The values of $E$ for which Eq.~(\ref{eqn:right_boundary_condition})
is satisfied exist in a countable set. These are the {\em physical\/}
eigenvalues of $E$.

In the case that $n=1$, numerically finding the value of $E$ for which
Eq.~(\ref{eqn:right_boundary_condition}) holds is accomplished via the
use of the {\it nodal theorem\/} of Sturm-Liouville systems. This
theorem has been generalized to systems for which $n\geq 1$
\cite{amann1995nodal}. The generalized nodal theorem allows one to
find the physical eigenvalues of $E$ in
Eq.~(\ref{eqn:johnson_schrod_U}) by using $\det \Umat(r)$, without
explicitly monitoring the functional solution $\Umat(r)$, to check
that Eq.~(\ref{eqn:right_boundary_condition}) is satisfied.

If $n=1$, one is assured that all of the nodes of $\Umat(r) \to u(r)$
are located in the {\em classically allowed\/} region. Hence, it is
important to know where the classical turning points are located. For
$n>1$, one has multiple potentials, and hence regions for which some
potentials may be in their classically allowed regions, while others
may be in their classically forbidden regions.

To make the notion of a ``classical turning point'' well defined in
the coupled case, one can solve for the roots of the eigenvalues of
$-\mathbf{Q}(r)$ as functions of $r$. For the case where each
potential has only one classically allowed region, the inner classical
turning point is defined as the innermost root of any of the
eigenvalues of $-\mathbf{Q}(r)$, and the outer classical turning point
is defined as the outermost root of any of the eigenvalues of
$-\mathbf{Q}(r)$. This definition of the classical turning points
ensures that no nodes are missed in the counting procedure due to
having skipped over a portion of the classically allowed region of one
of the potentials. Moreover, such a definition is established through
multichannel generalizations of the WKB approximation
\cite{johnson1973generalized}.

To find the desired value energy eigenvalue $E_{N}$ for which the
solution $\Umat_{N}$ has $N$ nodes, one first chooses a window of $E$
values that is bounded below by $E_{\textrm{low}}$, bounded above by
$E_{\textrm{high}}$, and that contains $E_N$. One then counts the
number of nodes $\Ns$ of $\det\Umat(r)$ in the classically allowed
region by numerically integrating Eq.~(\ref{eqn:johnson_schrod_U}) at
$E=E_{\textrm{mid}}=(E_{\textrm{low}}+E_{\textrm{high}})/2$. Once the
outer bound has been reached, one then sets
$E_{\textrm{low}}=E_{\textrm{mid}}$ if $\Ns\leq N$, or sets
$E_{\textrm{high}}=E_{\textrm{mid}}$ if $\Ns> N$. This procedure is
repeated until $E_{\textrm{high}}-E_{\textrm{low}}$ meets a
pre-specified tolerance.

\subsection{Renormalized Numerov Integration Procedure}

We now derive the renormalized Numerov method of
Ref. \cite{Johnson:1978}. Discretize the radial coordinate $r$ such
that the initial value $r_0$ is located to the left of the inner
classical turning point and is sufficiently close to the origin, while
the final value $r_f$ is located sufficiently to the right of the
outer classical turning point. Moreover, choose the grid to be
uniformly spaced, with $r_{i+1}-r_{i}=h$ for all
$i\in\{0,1,2,...,f\}$, and $h$ some arbitrarily small real
number. Moreover, let $\Umat(r_{i})\equiv\Umat_{i}$.

Numerical integration of Eq.~(\ref{eqn:johnson_schrod_U}) can be
accomplished by a coupled generalization of the Numerov recurrence
relation:
\begin{equation}\label{eqn:numerov}
    \left(\mathbb{I}-\Tmat_{i+2}\right)\Umat_{i+2}-\left(2\mathbb{I}+
    10\Tmat_{i+1}\right)\Umat_{i+1}+\left(\mathbb{I}-\Tmat_{i}\right)
    \Umat_{i}=\textbf{0} \, ,
\end{equation}
where
\begin{equation}
    \Tmat_i \equiv -\frac{h^2}{12}\mathbf{Q}(r_i) \, .
\end{equation}
The recurrence relation of the renormalized Numerov scheme results
from three substitutions:
\begin{eqnarray}\label{eqn:R_mat} 
    \Fmat_{i} & \equiv & \left( \mathbb{I}-\Tmat_{i} \right)
    \Umat_{i} \, , \nonumber \\
    \Hmat_i & \equiv & \left( 2\mathbb{I}+10\Tmat_{i} \right) \left(
    \mathbb{I}-\Tmat_{i} \right)^{-1} \, , \nonumber \\
    \Rmat_{i} & \equiv & \Fmat_{i+1} \Fmat^{-1}_{i} \, .
\end{eqnarray}
The first two substitutions turn Eq.~(\ref{eqn:numerov}) into
\begin{equation}\label{eqn:pre_renormal}
    \Fmat_{i+2}-\Hmat_{i+1}\Fmat_{i+1}+\Fmat_{i}=0 \, ,
\end{equation}
and the last substitution renormalizes Eq.~(\ref{eqn:pre_renormal}) as
\begin{equation}\label{eqn:renormalized_numerov}
    \Rmat_{i+1}=\Hmat_{i+1}-\Rmat^{-1}_{i} \, .
\end{equation}

Equation~(\ref{eqn:renormalized_numerov}) provides a stable and
efficient method for propagating $\Umat_{i}$ in regions where the
entries of $\mathbf{Q}$ diverge to $\pm\infty$, since $\Rmat_{i}\sim
\mathbf{Q}_{i+1}\mathbf{Q}^{-1}_{i}$. In most situations, one can
choose $\Rmat_{0}^{-1}=\mathbf{0}$; however, this choice can present
complications \cite{Johnson:1978}. We instead set $\Rmat^{-1}_0$ to be
\begin{equation}
       \Rmat^{-1}_0 = \left(\mathbb{I}-\Tmat_{0}\right)
       \left(2\mathbb{I}+10\Tmat_{0}\right)^{-1}=\Hmat_0^{-1} \, .
\end{equation}
If $r_0=h$, then this choice is equivalent to supposing that
$\Rmat_{-1}^{-1}=\mathbf{0}$ precisely at the origin, $r=r_{-1}=0$.

We now describe a method for counting the number of nodes along the
integration without explicitly monitoring $\det \Umat_{i}$ at each
integration step. First suppose that there is {\em only one\/} node of
$\det\Umat$ between $r_{i+1}$ and $r_i$. From the last of
Eq.~(\ref{eqn:R_mat}), one has that
\begin{equation}
    \det\Rmat_{i+1}=\frac{\det(\mathbb{I}-\Tmat_{i+1})}
{\det(\mathbb{I}-\Tmat_{i})}
    \times \frac{\det\Umat_{i+1}}{\det\Umat_{i}} \, .
\end{equation}
Since $\det(\mathbb{I}-\Tmat_{i+1})>0$ for any $r_{i}$ in the
classically allowed region, one has that $\det\Rmat_{i+1}<0$, since
$\det\Umat$ encountering one node between $r_{i+1}$ and $r_{i}$
implies that $\det \Umat_{i}$ and $\det \Umat_{i+1}$ have opposite
signs. It follows that one may monitor $\det \Rmat_{i}$ at each
integration step to count the number of nodes. Such a method is
effective if only one node exists between each pair of grid points, as
the existence of even numbers of nodes between any two grid points
(due to degeneracies, for example) implies that $\det \Umat_{i+1}$ and
$\det \Umat_{i}$ have the same sign, and hence $\det\Rmat_{i+1}>0$.

One can avoid missing nodes by instead monitoring the individual
eigenvalues of $\Rmat_{i}$ at each integration point. If $\det
\Rmat_{i}<0$, then there is an odd number of negative
eigenvalues. However, if $\det \Rmat_{i}>0$, there is an even number
of negative eigenvalues, and an incorrect node count
occurs. Therefore, if one instead monitors each time one of the
eigenvalues of $\Rmat_{i}$ changes sign, there is no danger of missing
a node. This procedure is equivalent to monitoring the signature of
$\Rmat_{i}$, and one can therefore equivalently count the number of
times one of the diagonal entries of the $U$ matrix in an
$LU$-decomposition of $\Rmat_{i}$ changes sign.

\vspace{1em}
\subsection{Calculation of Expectation Values}

From nondegenerate perturbation theory, one learns that if the
Hamiltonian $H$ splits into some reference Hamiltonian $H^{(0)}$ and a
small perturbation $\epsilon H'$, then the energy eigenvalues can be
calculated perturbatively. To first order in $\epsilon$, this
perturbative expansion is:
\begin{equation}\label{eqn:perturbation_theory}
    E'_{N}=E^{(0)}_{N}+\epsilon \langle N|H'|N\rangle + \mathcal{O}
    \left(\epsilon^2 \right) \, .
\end{equation}
The expectation value of $H'$ is then found as the limit
\begin{equation}\label{eqn:expectation_value}
    \langle N|H'|N\rangle=\lim_{\epsilon\rightarrow 0}
    \frac{E_{N}'-E^{(0)}_{N}}{\epsilon} \, .
\end{equation}

While this result when $H'$ is specifically a Hamiltonian perturbation
is the conventional Feynman-Hellmann theorem, it remains true for any
Hermitian operator $H'$. Therefore, Eq.~(\ref{eqn:expectation_value})
provides a method~\cite{Hutson:1994} for numerically calculating the
expectation values of arbitrary operators using just the energy
eigenvalues of the original problem and the energy eigenvalues of
Eq.~(\ref{eqn:perturbation_theory}), so long as these energies are
non-degenerate. This observation is very powerful, because it means
that one does not need to numerically compute the wave functions of
Eq.~(\ref{eqn:johnson_schrod_U}) in the calculation of expectation
values.

\bibliographystyle{apsrev4-1}
\bibliography{diquark}

\begin{thebibliography}{57}%
\makeatletter
\providecommand \@ifxundefined [1]{%
 \@ifx{#1\undefined}
}%
\providecommand \@ifnum [1]{%
 \ifnum #1\expandafter \@firstoftwo
 \else \expandafter \@secondoftwo
 \fi
}%
\providecommand \@ifx [1]{%
 \ifx #1\expandafter \@firstoftwo
 \else \expandafter \@secondoftwo
 \fi
}%
\providecommand \natexlab [1]{#1}%
\providecommand \enquote  [1]{``#1''}%
\providecommand \bibnamefont  [1]{#1}%
\providecommand \bibfnamefont [1]{#1}%
\providecommand \citenamefont [1]{#1}%
\providecommand \href@noop [0]{\@secondoftwo}%
\providecommand \href [0]{\begingroup \@sanitize@url \@href}%
\providecommand \@href[1]{\@@startlink{#1}\@@href}%
\providecommand \@@href[1]{\endgroup#1\@@endlink}%
\providecommand \@sanitize@url [0]{\catcode `\\12\catcode `\$12\catcode
  `\&12\catcode `\#12\catcode `\^12\catcode `\_12\catcode `\%12\relax}%
\providecommand \@@startlink[1]{}%
\providecommand \@@endlink[0]{}%
\providecommand \url  [0]{\begingroup\@sanitize@url \@url }%
\providecommand \@url [1]{\endgroup\@href {#1}{\urlprefix }}%
\providecommand \urlprefix  [0]{URL }%
\providecommand \Eprint [0]{\href }%
\providecommand \doibase [0]{http://dx.doi.org/}%
\providecommand \selectlanguage [0]{\@gobble}%
\providecommand \bibinfo  [0]{\@secondoftwo}%
\providecommand \bibfield  [0]{\@secondoftwo}%
\providecommand \translation [1]{[#1]}%
\providecommand \BibitemOpen [0]{}%
\providecommand \bibitemStop [0]{}%
\providecommand \bibitemNoStop [0]{.\EOS\space}%
\providecommand \EOS [0]{\spacefactor3000\relax}%
\providecommand \BibitemShut  [1]{\csname bibitem#1\endcsname}%
\let\auto@bib@innerbib\@empty
\bibitem [{\citenamefont {Lebed}\ \emph {et~al.}(2017)\citenamefont {Lebed},
  \citenamefont {Mitchell},\ and\ \citenamefont {Swanson}}]{Lebed:2016hpi}%
  \BibitemOpen
  \bibfield  {author} {\bibinfo {author} {\bibfnamefont {R.}~\bibnamefont
  {Lebed}}, \bibinfo {author} {\bibfnamefont {R.}~\bibnamefont {Mitchell}}, \
  and\ \bibinfo {author} {\bibfnamefont {E.}~\bibnamefont {Swanson}},\ }\href
  {\doibase 10.1016/j.ppnp.2016.11.003} {\bibfield  {journal} {\bibinfo
  {journal} {Prog.\ Part.\ Nucl.\ Phys.}\ }\textbf {\bibinfo {volume} {{\bf
  93}}},\ \bibinfo {pages} {143} (\bibinfo {year} {2017})},\ \Eprint
  {http://arxiv.org/abs/1610.04528} {arXiv:1610.04528 [hep-ph]} \BibitemShut
  {NoStop}%
\bibitem [{\citenamefont {Chen}\ \emph {et~al.}(2016)\citenamefont {Chen},
  \citenamefont {Chen}, \citenamefont {Liu},\ and\ \citenamefont
  {Zhu}}]{Chen:2016qju}%
  \BibitemOpen
  \bibfield  {author} {\bibinfo {author} {\bibfnamefont {H.-X.}\ \bibnamefont
  {Chen}}, \bibinfo {author} {\bibfnamefont {W.}~\bibnamefont {Chen}}, \bibinfo
  {author} {\bibfnamefont {X.}~\bibnamefont {Liu}}, \ and\ \bibinfo {author}
  {\bibfnamefont {S.-L.}\ \bibnamefont {Zhu}},\ }\href {\doibase
  10.1016/j.physrep.2016.05.004} {\bibfield  {journal} {\bibinfo  {journal}
  {Phys.\ Rept.}\ }\textbf {\bibinfo {volume} {{\bf 639}}},\ \bibinfo {pages}
  {1} (\bibinfo {year} {2016})},\ \Eprint {http://arxiv.org/abs/1601.02092}
  {arXiv:1601.02092 [hep-ph]} \BibitemShut {NoStop}%
\bibitem [{\citenamefont {Hosaka}\ \emph {et~al.}(2016)\citenamefont {Hosaka},
  \citenamefont {Iijima}, \citenamefont {Miyabayashi}, \citenamefont {Sakai},\
  and\ \citenamefont {Yasui}}]{Hosaka:2016pey}%
  \BibitemOpen
  \bibfield  {author} {\bibinfo {author} {\bibfnamefont {A.}~\bibnamefont
  {Hosaka}}, \bibinfo {author} {\bibfnamefont {T.}~\bibnamefont {Iijima}},
  \bibinfo {author} {\bibfnamefont {K.}~\bibnamefont {Miyabayashi}}, \bibinfo
  {author} {\bibfnamefont {Y.}~\bibnamefont {Sakai}}, \ and\ \bibinfo {author}
  {\bibfnamefont {S.}~\bibnamefont {Yasui}},\ }\href {\doibase
  10.1093/ptep/ptw045} {\bibfield  {journal} {\bibinfo  {journal} {Prog.\
  Theor.\ Exp.\ Phys.}\ }\textbf {\bibinfo {volume} {{\bf 2016}}},\ \bibinfo
  {pages} {062C01} (\bibinfo {year} {2016})},\ \Eprint
  {http://arxiv.org/abs/1603.09229} {arXiv:1603.09229 [hep-ph]} \BibitemShut
  {NoStop}%
\bibitem [{\citenamefont {Esposito}\ \emph {et~al.}(2016)\citenamefont
  {Esposito}, \citenamefont {Pilloni},\ and\ \citenamefont
  {Polosa}}]{Esposito:2016noz}%
  \BibitemOpen
  \bibfield  {author} {\bibinfo {author} {\bibfnamefont {A.}~\bibnamefont
  {Esposito}}, \bibinfo {author} {\bibfnamefont {A.}~\bibnamefont {Pilloni}}, \
  and\ \bibinfo {author} {\bibfnamefont {A.}~\bibnamefont {Polosa}},\ }\href
  {\doibase 10.1016/j.physrep.2016.11.002} {\bibfield  {journal} {\bibinfo
  {journal} {Phys.\ Rept.}\ }\textbf {\bibinfo {volume} {{\bf 668}}},\ \bibinfo
  {pages} {1} (\bibinfo {year} {2016})},\ \Eprint
  {http://arxiv.org/abs/1611.07920} {arXiv:1611.07920 [hep-ph]} \BibitemShut
  {NoStop}%
\bibitem [{\citenamefont {Guo}\ \emph {et~al.}(2018)\citenamefont {Guo},
  \citenamefont {Hanhart}, \citenamefont {Mei{\ss}ner}, \citenamefont {Wang},
  \citenamefont {Zhao},\ and\ \citenamefont {Zou}}]{Guo:2017jvc}%
  \BibitemOpen
  \bibfield  {author} {\bibinfo {author} {\bibfnamefont {F.-K.}\ \bibnamefont
  {Guo}}, \bibinfo {author} {\bibfnamefont {C.}~\bibnamefont {Hanhart}},
  \bibinfo {author} {\bibfnamefont {U.-G.}\ \bibnamefont {Mei{\ss}ner}},
  \bibinfo {author} {\bibfnamefont {Q.}~\bibnamefont {Wang}}, \bibinfo {author}
  {\bibfnamefont {Q.}~\bibnamefont {Zhao}}, \ and\ \bibinfo {author}
  {\bibfnamefont {B.-S.}\ \bibnamefont {Zou}},\ }\href {\doibase
  10.1103/RevModPhys.90.015004} {\bibfield  {journal} {\bibinfo  {journal}
  {Rev.\ Mod.\ Phys.}\ }\textbf {\bibinfo {volume} {{\bf 90}}},\ \bibinfo
  {pages} {015004} (\bibinfo {year} {2018})},\ \Eprint
  {http://arxiv.org/abs/1705.00141} {arXiv:1705.00141 [hep-ph]} \BibitemShut
  {NoStop}%
\bibitem [{\citenamefont {Ali}\ \emph {et~al.}(2017)\citenamefont {Ali},
  \citenamefont {Lange},\ and\ \citenamefont {Stone}}]{Ali:2017jda}%
  \BibitemOpen
  \bibfield  {author} {\bibinfo {author} {\bibfnamefont {A.}~\bibnamefont
  {Ali}}, \bibinfo {author} {\bibfnamefont {J.}~\bibnamefont {Lange}}, \ and\
  \bibinfo {author} {\bibfnamefont {S.}~\bibnamefont {Stone}},\ }\href
  {\doibase 10.1016/j.ppnp.2017.08.003} {\bibfield  {journal} {\bibinfo
  {journal} {Prog.\ Part.\ Nucl.\ Phys.}\ }\textbf {\bibinfo {volume} {{\bf
  97}}},\ \bibinfo {pages} {123} (\bibinfo {year} {2017})},\ \Eprint
  {http://arxiv.org/abs/1706.00610} {arXiv:1706.00610 [hep-ph]} \BibitemShut
  {NoStop}%
\bibitem [{\citenamefont {Olsen}\ \emph {et~al.}(2018)\citenamefont {Olsen},
  \citenamefont {Skwarnicki},\ and\ \citenamefont {Zieminska}}]{Olsen:2017bmm}%
  \BibitemOpen
  \bibfield  {author} {\bibinfo {author} {\bibfnamefont {S.}~\bibnamefont
  {Olsen}}, \bibinfo {author} {\bibfnamefont {T.}~\bibnamefont {Skwarnicki}}, \
  and\ \bibinfo {author} {\bibfnamefont {D.}~\bibnamefont {Zieminska}},\ }\href
  {\doibase 10.1103/RevModPhys.90.015003} {\bibfield  {journal} {\bibinfo
  {journal} {Rev.\ Mod.\ Phys.}\ }\textbf {\bibinfo {volume} {{\bf 90}}},\
  \bibinfo {pages} {015003} (\bibinfo {year} {2018})},\ \Eprint
  {http://arxiv.org/abs/1708.04012} {arXiv:1708.04012 [hep-ph]} \BibitemShut
  {NoStop}%
\bibitem [{\citenamefont {Karliner}\ \emph {et~al.}(2018)\citenamefont
  {Karliner}, \citenamefont {Rosner},\ and\ \citenamefont
  {Skwarnicki}}]{Karliner:2017qhf}%
  \BibitemOpen
  \bibfield  {author} {\bibinfo {author} {\bibfnamefont {M.}~\bibnamefont
  {Karliner}}, \bibinfo {author} {\bibfnamefont {J.}~\bibnamefont {Rosner}}, \
  and\ \bibinfo {author} {\bibfnamefont {T.}~\bibnamefont {Skwarnicki}},\
  }\href {\doibase 10.1146/annurev-nucl-101917-020902} {\bibfield  {journal}
  {\bibinfo  {journal} {Ann.\ Rev.\ Nucl.\ Part.\ Sci.}\ }\textbf {\bibinfo
  {volume} {{\bf 68}}},\ \bibinfo {pages} {17} (\bibinfo {year} {2018})},\
  \Eprint {http://arxiv.org/abs/1711.10626} {arXiv:1711.10626 [hep-ph]}
  \BibitemShut {NoStop}%
\bibitem [{\citenamefont {Yuan}(2018)}]{Yuan:2018inv}%
  \BibitemOpen
  \bibfield  {author} {\bibinfo {author} {\bibfnamefont {C.-Z.}\ \bibnamefont
  {Yuan}},\ }\href {\doibase 10.1142/S0217751X18300181} {\bibfield  {journal}
  {\bibinfo  {journal} {Int.\ J. Mod.\ Phys.}\ }\textbf {\bibinfo {volume}
  {{\bf A33}}},\ \bibinfo {pages} {1830018} (\bibinfo {year} {2018})},\ \Eprint
  {http://arxiv.org/abs/1808.01570} {arXiv:1808.01570 [hep-ex]} \BibitemShut
  {NoStop}%
\bibitem [{\citenamefont {Maiani}\ \emph {et~al.}(2005)\citenamefont {Maiani},
  \citenamefont {Piccinini}, \citenamefont {Polosa},\ and\ \citenamefont
  {Riquer}}]{Maiani:2004vq}%
  \BibitemOpen
  \bibfield  {author} {\bibinfo {author} {\bibfnamefont {L.}~\bibnamefont
  {Maiani}}, \bibinfo {author} {\bibfnamefont {F.}~\bibnamefont {Piccinini}},
  \bibinfo {author} {\bibfnamefont {A.}~\bibnamefont {Polosa}}, \ and\ \bibinfo
  {author} {\bibfnamefont {V.}~\bibnamefont {Riquer}},\ }\href {\doibase
  10.1103/PhysRevD.71.014028} {\bibfield  {journal} {\bibinfo  {journal}
  {Phys.\ Rev.}\ }\textbf {\bibinfo {volume} {D {\bf 71}}},\ \bibinfo {pages}
  {014028} (\bibinfo {year} {2005})},\ \Eprint
  {http://arxiv.org/abs/hep-ph/0412098} {arXiv:hep-ph/0412098 [hep-ph]}
  \BibitemShut {NoStop}%
\bibitem [{\citenamefont {Maiani}\ \emph {et~al.}(2014)\citenamefont {Maiani},
  \citenamefont {Piccinini}, \citenamefont {Polosa},\ and\ \citenamefont
  {Riquer}}]{Maiani:2014aja}%
  \BibitemOpen
  \bibfield  {author} {\bibinfo {author} {\bibfnamefont {L.}~\bibnamefont
  {Maiani}}, \bibinfo {author} {\bibfnamefont {F.}~\bibnamefont {Piccinini}},
  \bibinfo {author} {\bibfnamefont {A.}~\bibnamefont {Polosa}}, \ and\ \bibinfo
  {author} {\bibfnamefont {V.}~\bibnamefont {Riquer}},\ }\href {\doibase
  10.1103/PhysRevD.89.114010} {\bibfield  {journal} {\bibinfo  {journal}
  {Phys.\ Rev.}\ }\textbf {\bibinfo {volume} {D {\bf 89}}},\ \bibinfo {pages}
  {114010} (\bibinfo {year} {2014})},\ \Eprint {http://arxiv.org/abs/1405.1551}
  {arXiv:1405.1551 [hep-ph]} \BibitemShut {NoStop}%
\bibitem [{\citenamefont {Anselmino}\ \emph {et~al.}(1993)\citenamefont
  {Anselmino}, \citenamefont {Predazzi}, \citenamefont {Ekelin}, \citenamefont
  {Fredriksson},\ and\ \citenamefont {Lichtenberg}}]{Anselmino:1992vg}%
  \BibitemOpen
  \bibfield  {author} {\bibinfo {author} {\bibfnamefont {M.}~\bibnamefont
  {Anselmino}}, \bibinfo {author} {\bibfnamefont {E.}~\bibnamefont {Predazzi}},
  \bibinfo {author} {\bibfnamefont {S.}~\bibnamefont {Ekelin}}, \bibinfo
  {author} {\bibfnamefont {S.}~\bibnamefont {Fredriksson}}, \ and\ \bibinfo
  {author} {\bibfnamefont {D.}~\bibnamefont {Lichtenberg}},\ }\href {\doibase
  10.1103/RevModPhys.65.1199} {\bibfield  {journal} {\bibinfo  {journal} {Rev.\
  Mod.\ Phys.}\ }\textbf {\bibinfo {volume} {{\bf 65}}},\ \bibinfo {pages}
  {1199} (\bibinfo {year} {1993})}\BibitemShut {NoStop}%
\bibitem [{\citenamefont {Brodsky}\ \emph {et~al.}(2014)\citenamefont
  {Brodsky}, \citenamefont {Hwang},\ and\ \citenamefont
  {Lebed}}]{Brodsky:2014xia}%
  \BibitemOpen
  \bibfield  {author} {\bibinfo {author} {\bibfnamefont {S.}~\bibnamefont
  {Brodsky}}, \bibinfo {author} {\bibfnamefont {D.}~\bibnamefont {Hwang}}, \
  and\ \bibinfo {author} {\bibfnamefont {R.}~\bibnamefont {Lebed}},\ }\href
  {\doibase 10.1103/PhysRevLett.113.112001} {\bibfield  {journal} {\bibinfo
  {journal} {Phys.\ Rev.\ Lett.}\ }\textbf {\bibinfo {volume} {{\bf 113}}},\
  \bibinfo {pages} {112001} (\bibinfo {year} {2014})},\ \Eprint
  {http://arxiv.org/abs/1406.7281} {arXiv:1406.7281 [hep-ph]} \BibitemShut
  {NoStop}%
\bibitem [{\citenamefont {Brodsky}\ and\ \citenamefont
  {Lebed}(2015)}]{Brodsky:2015wza}%
  \BibitemOpen
  \bibfield  {author} {\bibinfo {author} {\bibfnamefont {S.}~\bibnamefont
  {Brodsky}}\ and\ \bibinfo {author} {\bibfnamefont {R.}~\bibnamefont
  {Lebed}},\ }\href {\doibase 10.1103/PhysRevD.91.114025} {\bibfield  {journal}
  {\bibinfo  {journal} {Phys.\ Rev.}\ }\textbf {\bibinfo {volume} {D {\bf
  91}}},\ \bibinfo {pages} {114025} (\bibinfo {year} {2015})},\ \Eprint
  {http://arxiv.org/abs/1505.00803} {arXiv:1505.00803 [hep-ph]} \BibitemShut
  {NoStop}%
\bibitem [{\citenamefont {Lebed}(2015)}]{Lebed:2015tna}%
  \BibitemOpen
  \bibfield  {author} {\bibinfo {author} {\bibfnamefont {R.}~\bibnamefont
  {Lebed}},\ }\href {\doibase 10.1016/j.physletb.2015.08.032} {\bibfield
  {journal} {\bibinfo  {journal} {Phys.\ Lett.}\ }\textbf {\bibinfo {volume} {B
  {\bf 749}}},\ \bibinfo {pages} {454} (\bibinfo {year} {2015})},\ \Eprint
  {http://arxiv.org/abs/1507.05867} {arXiv:1507.05867 [hep-ph]} \BibitemShut
  {NoStop}%
\bibitem [{\citenamefont {Lebed}(2017)}]{Lebed:2017min}%
  \BibitemOpen
  \bibfield  {author} {\bibinfo {author} {\bibfnamefont {R.}~\bibnamefont
  {Lebed}},\ }\href {\doibase 10.1103/PhysRevD.96.116003} {\bibfield  {journal}
  {\bibinfo  {journal} {Phys.\ Rev.}\ }\textbf {\bibinfo {volume} {D {\bf
  96}}},\ \bibinfo {pages} {116003} (\bibinfo {year} {2017})},\ \Eprint
  {http://arxiv.org/abs/1709.06097} {arXiv:1709.06097 [hep-ph]} \BibitemShut
  {NoStop}%
\bibitem [{\citenamefont {Eichten}\ \emph {et~al.}(1978)\citenamefont
  {Eichten}, \citenamefont {Gottfried}, \citenamefont {Kinoshita},
  \citenamefont {Lane},\ and\ \citenamefont {Yan}}]{Eichten:1978tg}%
  \BibitemOpen
  \bibfield  {author} {\bibinfo {author} {\bibfnamefont {E.}~\bibnamefont
  {Eichten}}, \bibinfo {author} {\bibfnamefont {K.}~\bibnamefont {Gottfried}},
  \bibinfo {author} {\bibfnamefont {T.}~\bibnamefont {Kinoshita}}, \bibinfo
  {author} {\bibfnamefont {K.}~\bibnamefont {Lane}}, \ and\ \bibinfo {author}
  {\bibfnamefont {T.-M.}\ \bibnamefont {Yan}},\ }\href {\doibase
  10.1103/PhysRevD.17.3090} {\bibfield  {journal} {\bibinfo  {journal} {Phys.\
  Rev.}\ }\textbf {\bibinfo {volume} {D {\bf 17}}},\ \bibinfo {pages} {3090}
  (\bibinfo {year} {1978})},\ \bibinfo {note} {[Erratum: Phys.\ Rev.\ D {\bf
  21}, 313 (1980)]}\BibitemShut {NoStop}%
\bibitem [{\citenamefont {Eichten}\ \emph {et~al.}(1980)\citenamefont
  {Eichten}, \citenamefont {Gottfried}, \citenamefont {Kinoshita},
  \citenamefont {Lane},\ and\ \citenamefont {Yan}}]{Eichten:1979ms}%
  \BibitemOpen
  \bibfield  {author} {\bibinfo {author} {\bibfnamefont {E.}~\bibnamefont
  {Eichten}}, \bibinfo {author} {\bibfnamefont {K.}~\bibnamefont {Gottfried}},
  \bibinfo {author} {\bibfnamefont {T.}~\bibnamefont {Kinoshita}}, \bibinfo
  {author} {\bibfnamefont {K.}~\bibnamefont {Lane}}, \ and\ \bibinfo {author}
  {\bibfnamefont {T.-M.}\ \bibnamefont {Yan}},\ }\href {\doibase
  10.1103/PhysRevD.21.203} {\bibfield  {journal} {\bibinfo  {journal} {Phys.\
  Rev.}\ }\textbf {\bibinfo {volume} {D {\bf 21}}},\ \bibinfo {pages} {203}
  (\bibinfo {year} {1980})}\BibitemShut {NoStop}%
\bibitem [{\citenamefont {Born}\ and\ \citenamefont
  {Oppenheimer}(1927)}]{Born:1927boa}%
  \BibitemOpen
  \bibfield  {author} {\bibinfo {author} {\bibfnamefont {M.}~\bibnamefont
  {Born}}\ and\ \bibinfo {author} {\bibfnamefont {R.}~\bibnamefont
  {Oppenheimer}},\ }\href {\doibase 10.1002/andp.19273892002} {\bibfield
  {journal} {\bibinfo  {journal} {Ann.\ der Phys.}\ }\textbf {\bibinfo {volume}
  {{\bf 389}}},\ \bibinfo {pages} {457} (\bibinfo {year} {1927})}\BibitemShut
  {NoStop}%
\bibitem [{\citenamefont {Griffiths}\ \emph {et~al.}(1983)\citenamefont
  {Griffiths}, \citenamefont {Michael},\ and\ \citenamefont
  {Rakow}}]{Griffiths:1983ah}%
  \BibitemOpen
  \bibfield  {author} {\bibinfo {author} {\bibfnamefont {L.}~\bibnamefont
  {Griffiths}}, \bibinfo {author} {\bibfnamefont {C.}~\bibnamefont {Michael}},
  \ and\ \bibinfo {author} {\bibfnamefont {P.}~\bibnamefont {Rakow}},\ }\href
  {\doibase 10.1016/0370-2693(83)90680-9} {\bibfield  {journal} {\bibinfo
  {journal} {Phys.\ Lett.}\ }\textbf {\bibinfo {volume} {{\bf 129B}}},\
  \bibinfo {pages} {351} (\bibinfo {year} {1983})}\BibitemShut {NoStop}%
\bibitem [{\citenamefont {Berwein}\ \emph {et~al.}(2015)\citenamefont
  {Berwein}, \citenamefont {Brambilla}, \citenamefont
  {Tarr{\'u}s~Castell{\`a}},\ and\ \citenamefont {Vairo}}]{Berwein:2015vca}%
  \BibitemOpen
  \bibfield  {author} {\bibinfo {author} {\bibfnamefont {M.}~\bibnamefont
  {Berwein}}, \bibinfo {author} {\bibfnamefont {N.}~\bibnamefont {Brambilla}},
  \bibinfo {author} {\bibfnamefont {J.}~\bibnamefont
  {Tarr{\'u}s~Castell{\`a}}}, \ and\ \bibinfo {author} {\bibfnamefont
  {A.}~\bibnamefont {Vairo}},\ }\href {\doibase 10.1103/PhysRevD.92.114019}
  {\bibfield  {journal} {\bibinfo  {journal} {Phys.\ Rev.}\ }\textbf {\bibinfo
  {volume} {D {\bf 92}}},\ \bibinfo {pages} {114019} (\bibinfo {year}
  {2015})},\ \Eprint {http://arxiv.org/abs/1510.04299} {arXiv:1510.04299
  [hep-ph]} \BibitemShut {NoStop}%
\bibitem [{\citenamefont {Lebed}\ and\ \citenamefont
  {Swanson}(2018)}]{Lebed:2017xih}%
  \BibitemOpen
  \bibfield  {author} {\bibinfo {author} {\bibfnamefont {R.}~\bibnamefont
  {Lebed}}\ and\ \bibinfo {author} {\bibfnamefont {E.}~\bibnamefont
  {Swanson}},\ }\href {\doibase 10.1007/s00601-018-1376-9} {\bibfield
  {journal} {\bibinfo  {journal} {Few Body Syst.}\ }\textbf {\bibinfo {volume}
  {{\bf 59}}},\ \bibinfo {pages} {53} (\bibinfo {year} {2018})},\ \Eprint
  {http://arxiv.org/abs/1708.02679} {arXiv:1708.02679 [hep-ph]} \BibitemShut
  {NoStop}%
\bibitem [{\citenamefont {Juge}\ \emph {et~al.}(1998)\citenamefont {Juge},
  \citenamefont {Kuti},\ and\ \citenamefont {Morningstar}}]{Juge:1997nc}%
  \BibitemOpen
  \bibfield  {author} {\bibinfo {author} {\bibfnamefont {K.}~\bibnamefont
  {Juge}}, \bibinfo {author} {\bibfnamefont {J.}~\bibnamefont {Kuti}}, \ and\
  \bibinfo {author} {\bibfnamefont {C.}~\bibnamefont {Morningstar}},\
  }\bibfield  {booktitle} {\emph {\bibinfo {booktitle} {{Contents of LAT97
  proceedings}}},\ }\href {\doibase 10.1016/S0920-5632(97)00759-7} {\bibfield
  {journal} {\bibinfo  {journal} {Nucl.\ Phys.\ Proc.\ Suppl.}\ }\textbf
  {\bibinfo {volume} {{\bf 63}}},\ \bibinfo {pages} {326} (\bibinfo {year}
  {1998})},\ \Eprint {http://arxiv.org/abs/hep-lat/9709131}
  {arXiv:hep-lat/9709131 [hep-lat]} \BibitemShut {NoStop}%
\bibitem [{\citenamefont {Juge}\ \emph {et~al.}(1999)\citenamefont {Juge},
  \citenamefont {Kuti},\ and\ \citenamefont {Morningstar}}]{Juge:1999ie}%
  \BibitemOpen
  \bibfield  {author} {\bibinfo {author} {\bibfnamefont {K.}~\bibnamefont
  {Juge}}, \bibinfo {author} {\bibfnamefont {J.}~\bibnamefont {Kuti}}, \ and\
  \bibinfo {author} {\bibfnamefont {C.}~\bibnamefont {Morningstar}},\ }\href
  {\doibase 10.1103/PhysRevLett.82.4400} {\bibfield  {journal} {\bibinfo
  {journal} {Phys.\ Rev.\ Lett.}\ }\textbf {\bibinfo {volume} {{\bf 82}}},\
  \bibinfo {pages} {4400} (\bibinfo {year} {1999})},\ \Eprint
  {http://arxiv.org/abs/hep-ph/9902336} {arXiv:hep-ph/9902336 [hep-ph]}
  \BibitemShut {NoStop}%
\bibitem [{\citenamefont {Juge}\ \emph {et~al.}(2003)\citenamefont {Juge},
  \citenamefont {Kuti},\ and\ \citenamefont {Morningstar}}]{Juge:2002br}%
  \BibitemOpen
  \bibfield  {author} {\bibinfo {author} {\bibfnamefont {K.}~\bibnamefont
  {Juge}}, \bibinfo {author} {\bibfnamefont {J.}~\bibnamefont {Kuti}}, \ and\
  \bibinfo {author} {\bibfnamefont {C.}~\bibnamefont {Morningstar}},\ }\href
  {\doibase 10.1103/PhysRevLett.90.161601} {\bibfield  {journal} {\bibinfo
  {journal} {Phys.\ Rev.\ Lett.}\ }\textbf {\bibinfo {volume} {{\bf 90}}},\
  \bibinfo {pages} {161601} (\bibinfo {year} {2003})},\ \Eprint
  {http://arxiv.org/abs/hep-lat/0207004} {arXiv:hep-lat/0207004 [hep-lat]}
  \BibitemShut {NoStop}%
\bibitem [{Mor()}]{Morningstar:2019}%
  \BibitemOpen
  \href@noop {} {}\bibinfo {howpublished}
  {\url{http://www.andrew.cmu.edu/user/cmorning/static_potentials/SU3_4D/greet.html}}\BibitemShut
  {NoStop}%
\bibitem [{\citenamefont {Capitani}\ \emph {et~al.}(2019)\citenamefont
  {Capitani}, \citenamefont {Philipsen}, \citenamefont {Reisinger},
  \citenamefont {Riehl},\ and\ \citenamefont {Wagner}}]{Capitani:2018rox}%
  \BibitemOpen
  \bibfield  {author} {\bibinfo {author} {\bibfnamefont {S.}~\bibnamefont
  {Capitani}}, \bibinfo {author} {\bibfnamefont {O.}~\bibnamefont {Philipsen}},
  \bibinfo {author} {\bibfnamefont {C.}~\bibnamefont {Reisinger}}, \bibinfo
  {author} {\bibfnamefont {C.}~\bibnamefont {Riehl}}, \ and\ \bibinfo {author}
  {\bibfnamefont {M.}~\bibnamefont {Wagner}},\ }\href {\doibase
  10.1103/PhysRevD.99.034502} {\bibfield  {journal} {\bibinfo  {journal}
  {Phys.\ Rev.}\ }\textbf {\bibinfo {volume} {D {\bf 99}}},\ \bibinfo {pages}
  {034502} (\bibinfo {year} {2019})},\ \Eprint
  {http://arxiv.org/abs/1811.11046} {arXiv:1811.11046 [hep-lat]} \BibitemShut
  {NoStop}%
\bibitem [{\citenamefont {Bali}\ and\ \citenamefont
  {Pineda}(2004)}]{Bali:2003jq}%
  \BibitemOpen
  \bibfield  {author} {\bibinfo {author} {\bibfnamefont {G.}~\bibnamefont
  {Bali}}\ and\ \bibinfo {author} {\bibfnamefont {A.}~\bibnamefont {Pineda}},\
  }\href {\doibase 10.1103/PhysRevD.69.094001} {\bibfield  {journal} {\bibinfo
  {journal} {Phys.\ Rev.}\ }\textbf {\bibinfo {volume} {D {\bf 69}}},\ \bibinfo
  {pages} {094001} (\bibinfo {year} {2004})},\ \Eprint
  {http://arxiv.org/abs/hep-ph/0310130} {arXiv:hep-ph/0310130 [hep-ph]}
  \BibitemShut {NoStop}%
\bibitem [{\citenamefont {Bicudo}\ \emph {et~al.}(2017)\citenamefont {Bicudo},
  \citenamefont {Cardoso}, \citenamefont {Oliveira},\ and\ \citenamefont
  {Silva}}]{Bicudo:2017usw}%
  \BibitemOpen
  \bibfield  {author} {\bibinfo {author} {\bibfnamefont {P.}~\bibnamefont
  {Bicudo}}, \bibinfo {author} {\bibfnamefont {M.}~\bibnamefont {Cardoso}},
  \bibinfo {author} {\bibfnamefont {O.}~\bibnamefont {Oliveira}}, \ and\
  \bibinfo {author} {\bibfnamefont {P.}~\bibnamefont {Silva}},\ }\href
  {\doibase 10.1103/PhysRevD.96.074508} {\bibfield  {journal} {\bibinfo
  {journal} {Phys.\ Rev.}\ }\textbf {\bibinfo {volume} {D {\bf 96}}},\ \bibinfo
  {pages} {074508} (\bibinfo {year} {2017})},\ \Eprint
  {http://arxiv.org/abs/1702.07789} {arXiv:1702.07789 [hep-lat]} \BibitemShut
  {NoStop}%
\bibitem [{\citenamefont {Braaten}(2013)}]{Braaten:2013boa}%
  \BibitemOpen
  \bibfield  {author} {\bibinfo {author} {\bibfnamefont {E.}~\bibnamefont
  {Braaten}},\ }\href {\doibase 10.1103/PhysRevLett.111.162003} {\bibfield
  {journal} {\bibinfo  {journal} {Phys.\ Rev.\ Lett.}\ }\textbf {\bibinfo
  {volume} {{\bf 111}}},\ \bibinfo {pages} {162003} (\bibinfo {year} {2013})},\
  \Eprint {http://arxiv.org/abs/1305.6905} {arXiv:1305.6905 [hep-ph]}
  \BibitemShut {NoStop}%
\bibitem [{\citenamefont {Braaten}\ \emph
  {et~al.}(2014{\natexlab{a}})\citenamefont {Braaten}, \citenamefont
  {Langmack},\ and\ \citenamefont {Smith}}]{Braaten:2014ita}%
  \BibitemOpen
  \bibfield  {author} {\bibinfo {author} {\bibfnamefont {E.}~\bibnamefont
  {Braaten}}, \bibinfo {author} {\bibfnamefont {C.}~\bibnamefont {Langmack}}, \
  and\ \bibinfo {author} {\bibfnamefont {D.}~\bibnamefont {Smith}},\ }\href
  {\doibase 10.1103/PhysRevLett.112.222001} {\bibfield  {journal} {\bibinfo
  {journal} {Phys.\ Rev.\ Lett.}\ }\textbf {\bibinfo {volume} {{\bf 112}}},\
  \bibinfo {pages} {222001} (\bibinfo {year} {2014}{\natexlab{a}})},\ \Eprint
  {http://arxiv.org/abs/1401.7351} {arXiv:1401.7351 [hep-ph]} \BibitemShut
  {NoStop}%
\bibitem [{\citenamefont {Braaten}\ \emph
  {et~al.}(2014{\natexlab{b}})\citenamefont {Braaten}, \citenamefont
  {Langmack},\ and\ \citenamefont {Smith}}]{Braaten:2014qka}%
  \BibitemOpen
  \bibfield  {author} {\bibinfo {author} {\bibfnamefont {E.}~\bibnamefont
  {Braaten}}, \bibinfo {author} {\bibfnamefont {C.}~\bibnamefont {Langmack}}, \
  and\ \bibinfo {author} {\bibfnamefont {D.}~\bibnamefont {Smith}},\ }\href
  {\doibase 10.1103/PhysRevD.90.014044} {\bibfield  {journal} {\bibinfo
  {journal} {Phys.\ Rev.}\ }\textbf {\bibinfo {volume} {D {\bf 90}}},\ \bibinfo
  {pages} {014044} (\bibinfo {year} {2014}{\natexlab{b}})},\ \Eprint
  {http://arxiv.org/abs/1402.0438} {arXiv:1402.0438 [hep-ph]} \BibitemShut
  {NoStop}%
\bibitem [{\citenamefont {Aubert}\ and\ \citenamefont {{\it et
  al.}}(2005)}]{Aubert:2004zr}%
  \BibitemOpen
  \bibfield  {author} {\bibinfo {author} {\bibfnamefont {B.}~\bibnamefont
  {Aubert}}\ and\ \bibinfo {author} {\bibnamefont {{\it et al.}}} (\bibinfo
  {collaboration} {BaBar Collaboration}),\ }\href {\doibase
  10.1103/PhysRevD.71.031501} {\bibfield  {journal} {\bibinfo  {journal}
  {Phys.\ Rev.}\ }\textbf {\bibinfo {volume} {D {\bf 71}}},\ \bibinfo {pages}
  {031501} (\bibinfo {year} {2005})},\ \Eprint
  {http://arxiv.org/abs/hep-ex/0412051} {arXiv:hep-ex/0412051 [hep-ex]}
  \BibitemShut {NoStop}%
\bibitem [{\citenamefont {Aaij}\ and\ \citenamefont {{\it et
  al.}}(2019)}]{Aaij:2019vzc}%
  \BibitemOpen
  \bibfield  {author} {\bibinfo {author} {\bibfnamefont {R.}~\bibnamefont
  {Aaij}}\ and\ \bibinfo {author} {\bibnamefont {{\it et al.}}} (\bibinfo
  {collaboration} {LHCb Collaboration}),\ }\href@noop {} {\  (\bibinfo {year}
  {2019})},\ \Eprint {http://arxiv.org/abs/1904.03947} {arXiv:1904.03947
  [hep-ex]} \BibitemShut {NoStop}%
\bibitem [{\citenamefont {Caswell}\ and\ \citenamefont
  {Lepage}(1986)}]{Caswell:1985ui}%
  \BibitemOpen
  \bibfield  {author} {\bibinfo {author} {\bibfnamefont {W.}~\bibnamefont
  {Caswell}}\ and\ \bibinfo {author} {\bibfnamefont {G.}~\bibnamefont
  {Lepage}},\ }\href {\doibase 10.1016/0370-2693(86)91297-9} {\bibfield
  {journal} {\bibinfo  {journal} {Phys.\ Lett.}\ }\textbf {\bibinfo {volume}
  {{\bf 167B}}},\ \bibinfo {pages} {437} (\bibinfo {year} {1986})}\BibitemShut
  {NoStop}%
\bibitem [{\citenamefont {Bodwin}\ \emph {et~al.}(1995)\citenamefont {Bodwin},
  \citenamefont {Braaten},\ and\ \citenamefont {Lepage}}]{Bodwin:1994jh}%
  \BibitemOpen
  \bibfield  {author} {\bibinfo {author} {\bibfnamefont {G.}~\bibnamefont
  {Bodwin}}, \bibinfo {author} {\bibfnamefont {E.}~\bibnamefont {Braaten}}, \
  and\ \bibinfo {author} {\bibfnamefont {G.}~\bibnamefont {Lepage}},\ }\href
  {\doibase 10.1103/PhysRevD.51.1125} {\bibfield  {journal} {\bibinfo
  {journal} {Phys.\ Rev.}\ }\textbf {\bibinfo {volume} {D {\bf 51}}},\ \bibinfo
  {pages} {1125} (\bibinfo {year} {1995})},\ \bibinfo {note} {[Erratum: Phys.\
  Rev.\ D {\bf 55},5853 (1997)]},\ \Eprint
  {http://arxiv.org/abs/hep-ph/9407339} {arXiv:hep-ph/9407339 [hep-ph]}
  \BibitemShut {NoStop}%
\bibitem [{\citenamefont {Pineda}\ and\ \citenamefont
  {Soto}(1998)}]{Pineda:1997bj}%
  \BibitemOpen
  \bibfield  {author} {\bibinfo {author} {\bibfnamefont {A.}~\bibnamefont
  {Pineda}}\ and\ \bibinfo {author} {\bibfnamefont {J.}~\bibnamefont {Soto}},\
  }\bibfield  {booktitle} {\emph {\bibinfo {booktitle} {{Quantum
  Chromodynamics. Proceedings, Conference, QCD'97, Montpellier, France, July
  3--9, 1997}}},\ }\href {\doibase 10.1016/S0920-5632(97)01102-X} {\bibfield
  {journal} {\bibinfo  {journal} {Nucl.\ Phys.\ Proc.\ Suppl.}\ }\textbf
  {\bibinfo {volume} {{\bf 64}}},\ \bibinfo {pages} {428} (\bibinfo {year}
  {1998})},\ \Eprint {http://arxiv.org/abs/hep-ph/9707481}
  {arXiv:hep-ph/9707481 [hep-ph]} \BibitemShut {NoStop}%
\bibitem [{\citenamefont {Brambilla}\ \emph {et~al.}(2000)\citenamefont
  {Brambilla}, \citenamefont {Pineda}, \citenamefont {Soto},\ and\
  \citenamefont {Vairo}}]{Brambilla:1999xf}%
  \BibitemOpen
  \bibfield  {author} {\bibinfo {author} {\bibfnamefont {N.}~\bibnamefont
  {Brambilla}}, \bibinfo {author} {\bibfnamefont {A.}~\bibnamefont {Pineda}},
  \bibinfo {author} {\bibfnamefont {J.}~\bibnamefont {Soto}}, \ and\ \bibinfo
  {author} {\bibfnamefont {A.}~\bibnamefont {Vairo}},\ }\href {\doibase
  10.1016/S0550-3213(99)00693-8} {\bibfield  {journal} {\bibinfo  {journal}
  {Nucl.\ Phys.}\ }\textbf {\bibinfo {volume} {{\bf B566}}},\ \bibinfo {pages}
  {275} (\bibinfo {year} {2000})},\ \Eprint
  {http://arxiv.org/abs/hep-ph/9907240} {arXiv:hep-ph/9907240 [hep-ph]}
  \BibitemShut {NoStop}%
\bibitem [{\citenamefont {Landau}\ and\ \citenamefont
  {Lifshitz}(1977)}]{Landau:1977}%
  \BibitemOpen
  \bibfield  {author} {\bibinfo {author} {\bibfnamefont {L.}~\bibnamefont
  {Landau}}\ and\ \bibinfo {author} {\bibfnamefont {E.}~\bibnamefont
  {Lifshitz}},\ }\href@noop {} {\emph {\bibinfo {title} {{\em Quantum
  Mechanics: Non-Relativistic Theory}}}},\ {\em Course of Theoretical Physics}\
  (\bibinfo  {publisher} {Pergamon Press},\ \bibinfo {address} {Oxford, U.K.},\
  \bibinfo {year} {1977})\BibitemShut {NoStop}%
\bibitem [{\citenamefont {Johnson}(1978)}]{Johnson:1978}%
  \BibitemOpen
  \bibfield  {author} {\bibinfo {author} {\bibfnamefont {B.}~\bibnamefont
  {Johnson}},\ }\href {\doibase 10.1063/1.436421} {\bibfield  {journal}
  {\bibinfo  {journal} {J. Chem.\ Phys.}\ }\textbf {\bibinfo {volume} {{\bf
  69}}},\ \bibinfo {pages} {4678} (\bibinfo {year} {1978})}\BibitemShut
  {NoStop}%
\bibitem [{\citenamefont {Hutson}(1994)}]{Hutson:1994}%
  \BibitemOpen
  \bibfield  {author} {\bibinfo {author} {\bibfnamefont {J.}~\bibnamefont
  {Hutson}},\ }\href {\doibase 10.1016/0010-4655(94)90200-3} {\bibfield
  {journal} {\bibinfo  {journal} {Comp.\ Phys.\ Comm.}\ }\textbf {\bibinfo
  {volume} {{\bf 84}}},\ \bibinfo {pages} {1} (\bibinfo {year}
  {1994})}\BibitemShut {NoStop}%
\bibitem [{\citenamefont {Barnes}\ \emph {et~al.}(2005)\citenamefont {Barnes},
  \citenamefont {Godfrey},\ and\ \citenamefont {Swanson}}]{Barnes:2005pb}%
  \BibitemOpen
  \bibfield  {author} {\bibinfo {author} {\bibfnamefont {T.}~\bibnamefont
  {Barnes}}, \bibinfo {author} {\bibfnamefont {S.}~\bibnamefont {Godfrey}}, \
  and\ \bibinfo {author} {\bibfnamefont {E.}~\bibnamefont {Swanson}},\ }\href
  {\doibase 10.1103/PhysRevD.72.054026} {\bibfield  {journal} {\bibinfo
  {journal} {Phys.\ Rev.}\ }\textbf {\bibinfo {volume} {D {\bf 72}}},\ \bibinfo
  {pages} {054026} (\bibinfo {year} {2005})},\ \Eprint
  {http://arxiv.org/abs/hep-ph/0505002} {arXiv:hep-ph/0505002 [hep-ph]}
  \BibitemShut {NoStop}%
\bibitem [{\citenamefont {Tanabashi}\ and\ \citenamefont {{\it et
  al.}}(2018)}]{Tanabashi:2018oca}%
  \BibitemOpen
  \bibfield  {author} {\bibinfo {author} {\bibfnamefont {M.}~\bibnamefont
  {Tanabashi}}\ and\ \bibinfo {author} {\bibnamefont {{\it et al.}}} (\bibinfo
  {collaboration} {Particle Data Group}),\ }\href {\doibase
  10.1103/PhysRevD.98.030001} {\bibfield  {journal} {\bibinfo  {journal}
  {Phys.\ Rev.}\ }\textbf {\bibinfo {volume} {D {\bf 98}}},\ \bibinfo {pages}
  {030001} (\bibinfo {year} {2018})}\BibitemShut {NoStop}%
\bibitem [{\citenamefont {Lebed}\ and\ \citenamefont
  {Polosa}(2016)}]{Lebed:2016yvr}%
  \BibitemOpen
  \bibfield  {author} {\bibinfo {author} {\bibfnamefont {R.}~\bibnamefont
  {Lebed}}\ and\ \bibinfo {author} {\bibfnamefont {A.}~\bibnamefont {Polosa}},\
  }\href {\doibase 10.1103/PhysRevD.93.094024} {\bibfield  {journal} {\bibinfo
  {journal} {Phys.\ Rev.}\ }\textbf {\bibinfo {volume} {D {\bf 93}}},\ \bibinfo
  {pages} {094024} (\bibinfo {year} {2016})},\ \Eprint
  {http://arxiv.org/abs/1602.08421} {arXiv:1602.08421 [hep-ph]} \BibitemShut
  {NoStop}%
\bibitem [{\citenamefont {Aaij}\ and\ \citenamefont {{\it et
  al.}}(2018)}]{Aaij:2018bla}%
  \BibitemOpen
  \bibfield  {author} {\bibinfo {author} {\bibfnamefont {R.}~\bibnamefont
  {Aaij}}\ and\ \bibinfo {author} {\bibnamefont {{\it et al.}}} (\bibinfo
  {collaboration} {LHCb Collaboration}),\ }\href {\doibase
  10.1140/epjc/s10052-018-6447-z} {\bibfield  {journal} {\bibinfo  {journal}
  {Eur.\ Phys.\ J.}\ }\textbf {\bibinfo {volume} {{\bf C78}}},\ \bibinfo
  {pages} {1019} (\bibinfo {year} {2018})},\ \Eprint
  {http://arxiv.org/abs/1809.07416} {arXiv:1809.07416 [hep-ex]} \BibitemShut
  {NoStop}%
\bibitem [{\citenamefont {Zhou}\ \emph {et~al.}(2015)\citenamefont {Zhou},
  \citenamefont {Xiao},\ and\ \citenamefont {Zhou}}]{Zhou:2015uva}%
  \BibitemOpen
  \bibfield  {author} {\bibinfo {author} {\bibfnamefont {Z.-Y.}\ \bibnamefont
  {Zhou}}, \bibinfo {author} {\bibfnamefont {Z.}~\bibnamefont {Xiao}}, \ and\
  \bibinfo {author} {\bibfnamefont {H.-Q.}\ \bibnamefont {Zhou}},\ }\href
  {\doibase 10.1103/PhysRevLett.115.022001} {\bibfield  {journal} {\bibinfo
  {journal} {Phys.\ Rev.\ Lett.}\ }\textbf {\bibinfo {volume} {115}},\ \bibinfo
  {pages} {022001} (\bibinfo {year} {2015})},\ \Eprint
  {http://arxiv.org/abs/1501.00879} {arXiv:1501.00879 [hep-ph]} \BibitemShut
  {NoStop}%
\bibitem [{\citenamefont {Chilikin}\ and\ \citenamefont {{\it et
  al.}}(2017)}]{Chilikin:2017evr}%
  \BibitemOpen
  \bibfield  {author} {\bibinfo {author} {\bibfnamefont {K.}~\bibnamefont
  {Chilikin}}\ and\ \bibinfo {author} {\bibnamefont {{\it et al.}}} (\bibinfo
  {collaboration} {Belle Collaboration}),\ }\href {\doibase
  10.1103/PhysRevD.95.112003} {\bibfield  {journal} {\bibinfo  {journal}
  {Phys.\ Rev.}\ }\textbf {\bibinfo {volume} {D {\bf 95}}},\ \bibinfo {pages}
  {112003} (\bibinfo {year} {2017})},\ \Eprint
  {http://arxiv.org/abs/1704.01872} {arXiv:1704.01872 [hep-ex]} \BibitemShut
  {NoStop}%
\bibitem [{\citenamefont {Ablikim}\ and\ \citenamefont {{\it et
  al.}}(2017{\natexlab{a}})}]{Ablikim:2017oaf}%
  \BibitemOpen
  \bibfield  {author} {\bibinfo {author} {\bibfnamefont {M.}~\bibnamefont
  {Ablikim}}\ and\ \bibinfo {author} {\bibnamefont {{\it et al.}}} (\bibinfo
  {collaboration} {BESIII Collaboration}),\ }\href {\doibase
  10.1103/PhysRevD.96.032004} {\bibfield  {journal} {\bibinfo  {journal}
  {Phys.\ Rev.}\ }\textbf {\bibinfo {volume} {D {\bf 96}}},\ \bibinfo {pages}
  {032004} (\bibinfo {year} {2017}{\natexlab{a}})},\ \bibinfo {note} {[Erratum:
  Phys.\ Rev.\ D {\bf 99}, 019903 (2019)]},\ \Eprint
  {http://arxiv.org/abs/1703.08787} {arXiv:1703.08787 [hep-ex]} \BibitemShut
  {NoStop}%
\bibitem [{\citenamefont {Ablikim}\ and\ \citenamefont {{\it et
  al.}}(2018)}]{Ablikim:2017aji}%
  \BibitemOpen
  \bibfield  {author} {\bibinfo {author} {\bibfnamefont {M.}~\bibnamefont
  {Ablikim}}\ and\ \bibinfo {author} {\bibnamefont {{\it et al.}}} (\bibinfo
  {collaboration} {BESIII Collaboration}),\ }\href {\doibase
  10.1103/PhysRevD.97.052001} {\bibfield  {journal} {\bibinfo  {journal}
  {Phys.\ Rev.}\ }\textbf {\bibinfo {volume} {D {\bf 97}}},\ \bibinfo {pages}
  {052001} (\bibinfo {year} {2018})},\ \Eprint
  {http://arxiv.org/abs/1710.10740} {arXiv:1710.10740 [hep-ex]} \BibitemShut
  {NoStop}%
\bibitem [{\citenamefont {Yuan}\ and\ \citenamefont {{\it et
  al.}}(2007)}]{Yuan:2007sj}%
  \BibitemOpen
  \bibfield  {author} {\bibinfo {author} {\bibfnamefont {C.-Z.}\ \bibnamefont
  {Yuan}}\ and\ \bibinfo {author} {\bibnamefont {{\it et al.}}} (\bibinfo
  {collaboration} {Belle Collaboration}),\ }\href {\doibase
  10.1103/PhysRevLett.99.182004} {\bibfield  {journal} {\bibinfo  {journal}
  {Phys.\ Rev.\ Lett.}\ }\textbf {\bibinfo {volume} {{\bf 99}}},\ \bibinfo
  {pages} {182004} (\bibinfo {year} {2007})},\ \Eprint
  {http://arxiv.org/abs/0707.2541} {arXiv:0707.2541 [hep-ex]} \BibitemShut
  {NoStop}%
\bibitem [{\citenamefont {Ablikim}\ and\ \citenamefont {{\it et
  al.}}(2017{\natexlab{b}})}]{BESIII:2016adj}%
  \BibitemOpen
  \bibfield  {author} {\bibinfo {author} {\bibfnamefont {M.}~\bibnamefont
  {Ablikim}}\ and\ \bibinfo {author} {\bibnamefont {{\it et al.}}} (\bibinfo
  {collaboration} {BESIII Collaboration}),\ }\href {\doibase
  10.1103/PhysRevLett.118.092002} {\bibfield  {journal} {\bibinfo  {journal}
  {Phys.\ Rev.\ Lett.}\ }\textbf {\bibinfo {volume} {{\bf 118}}},\ \bibinfo
  {pages} {092002} (\bibinfo {year} {2017}{\natexlab{b}})},\ \Eprint
  {http://arxiv.org/abs/1610.07044} {arXiv:1610.07044 [hep-ex]} \BibitemShut
  {NoStop}%
\bibitem [{\citenamefont {Bulava}\ \emph {et~al.}(2019)\citenamefont {Bulava},
  \citenamefont {H{\"o}rz}, \citenamefont {Knechtli}, \citenamefont {Koch},
  \citenamefont {Moir}, \citenamefont {Morningstar},\ and\ \citenamefont
  {Peardon}}]{Bulava:2019iut}%
  \BibitemOpen
  \bibfield  {author} {\bibinfo {author} {\bibfnamefont {J.}~\bibnamefont
  {Bulava}}, \bibinfo {author} {\bibfnamefont {B.}~\bibnamefont {H{\"o}rz}},
  \bibinfo {author} {\bibfnamefont {F.}~\bibnamefont {Knechtli}}, \bibinfo
  {author} {\bibfnamefont {V.}~\bibnamefont {Koch}}, \bibinfo {author}
  {\bibfnamefont {G.}~\bibnamefont {Moir}}, \bibinfo {author} {\bibfnamefont
  {C.}~\bibnamefont {Morningstar}}, \ and\ \bibinfo {author} {\bibfnamefont
  {M.}~\bibnamefont {Peardon}},\ }\href@noop {} {\  (\bibinfo {year} {2019})},\
  \Eprint {http://arxiv.org/abs/1902.04006} {arXiv:1902.04006 [hep-lat]}
  \BibitemShut {NoStop}%
\bibitem [{\citenamefont {Giron}\ \emph {et~al.}(2019)\citenamefont {Giron},
  \citenamefont {Lebed},\ and\ \citenamefont {Peterson}}]{Giron:2019}%
  \BibitemOpen
  \bibfield  {author} {\bibinfo {author} {\bibfnamefont {J.}~\bibnamefont
  {Giron}}, \bibinfo {author} {\bibfnamefont {R.}~\bibnamefont {Lebed}}, \ and\
  \bibinfo {author} {\bibfnamefont {C.}~\bibnamefont {Peterson}},\ }\href@noop
  {} {\  (\bibinfo {year} {2019})},\ \bibinfo {note} {in
  preparation}\BibitemShut {NoStop}%
\bibitem [{\citenamefont {Cleven}\ \emph {et~al.}(2015)\citenamefont {Cleven},
  \citenamefont {Guo}, \citenamefont {Hanhart}, \citenamefont {Wang},\ and\
  \citenamefont {Zhao}}]{Cleven:2015era}%
  \BibitemOpen
  \bibfield  {author} {\bibinfo {author} {\bibfnamefont {M.}~\bibnamefont
  {Cleven}}, \bibinfo {author} {\bibfnamefont {F.-K.}\ \bibnamefont {Guo}},
  \bibinfo {author} {\bibfnamefont {C.}~\bibnamefont {Hanhart}}, \bibinfo
  {author} {\bibfnamefont {Q.}~\bibnamefont {Wang}}, \ and\ \bibinfo {author}
  {\bibfnamefont {Q.}~\bibnamefont {Zhao}},\ }\href {\doibase
  10.1103/PhysRevD.92.014005} {\bibfield  {journal} {\bibinfo  {journal}
  {Phys.\ Rev.}\ }\textbf {\bibinfo {volume} {D {\bf 92}}},\ \bibinfo {pages}
  {014005} (\bibinfo {year} {2015})},\ \Eprint
  {http://arxiv.org/abs/1505.01771} {arXiv:1505.01771 [hep-ph]} \BibitemShut
  {NoStop}%
\bibitem [{\citenamefont {Alford}\ \emph {et~al.}(1999)\citenamefont {Alford},
  \citenamefont {Rajagopal},\ and\ \citenamefont {Wilczek}}]{Alford:1998mk}%
  \BibitemOpen
  \bibfield  {author} {\bibinfo {author} {\bibfnamefont {M.}~\bibnamefont
  {Alford}}, \bibinfo {author} {\bibfnamefont {K.}~\bibnamefont {Rajagopal}}, \
  and\ \bibinfo {author} {\bibfnamefont {F.}~\bibnamefont {Wilczek}},\ }\href
  {\doibase 10.1016/S0550-3213(98)00668-3} {\bibfield  {journal} {\bibinfo
  {journal} {Nucl.\ Phys.}\ }\textbf {\bibinfo {volume} {{\bf B537}}},\
  \bibinfo {pages} {443} (\bibinfo {year} {1999})},\ \Eprint
  {http://arxiv.org/abs/hep-ph/9804403} {arXiv:hep-ph/9804403 [hep-ph]}
  \BibitemShut {NoStop}%
\bibitem [{\citenamefont {Amann}\ and\ \citenamefont
  {Quittner}(1995)}]{amann1995nodal}%
  \BibitemOpen
  \bibfield  {author} {\bibinfo {author} {\bibfnamefont {H.}~\bibnamefont
  {Amann}}\ and\ \bibinfo {author} {\bibfnamefont {P.}~\bibnamefont
  {Quittner}},\ }\href@noop {} {\bibfield  {journal} {\bibinfo  {journal} {J.
  Math.\ Phys.}\ }\textbf {\bibinfo {volume} {{\bf 36}}},\ \bibinfo {pages}
  {4553} (\bibinfo {year} {1995})}\BibitemShut {NoStop}%
\bibitem [{\citenamefont {Johnson}(1973)}]{johnson1973generalized}%
  \BibitemOpen
  \bibfield  {author} {\bibinfo {author} {\bibfnamefont {B.}~\bibnamefont
  {Johnson}},\ }\href@noop {} {\bibfield  {journal} {\bibinfo  {journal}
  {Chem.\ Phys.}\ }\textbf {\bibinfo {volume} {{\bf 36}}},\ \bibinfo {pages}
  {381} (\bibinfo {year} {1973})}\BibitemShut {NoStop}%
\end{thebibliography}%


%
\end{document}